**Parity-time Symmetry in Non-Hermitian Complex Optical Media**

*Samit Kumar Gupta, Yi Zou, Xue-Yi Zhu, Ming-Hui Lu,[*] Lijian Zhang,[*] Xiao-Ping Liu,[*] and Yan-Feng Chen[*]*

Dr. S. K. Gupta, Dr. Y. Zou, Mr. X. -Y. Zhu
National Laboratory of Solid-State Microstructures, College of Engineering and Applied Sciences
Nanjing University
Nanjing 210093, P. R. China

Prof. M. -H. Lu, Prof. L. Zhang, Prof. X. -P. Liu, Prof. Y. -F. Chen
National Laboratory of Solid-State Microstructures, College of Engineering and Applied Sciences and Collaborative Innovation Center of Advanced Microstructures
Nanjing University
Nanjing 210093, P. R. China
Emails: luminghui@nju.edu.cn, lijian.zhang@nju.edu.cn, xpliu@nju.edu.cn, yfchen@nju.edu.cn

Abstract

The explorations of the quantum-inspired symmetries in optical and photonic systems have witnessed immense research interests both fundamentally and technologically in a wide range of subjects of physics and engineering. One of the principal emerging fields in this context is non-Hermitian physics based on parity-time symmetry, originally proposed in the studies pertaining to quantum mechanics and quantum field theory, recently ramified into diverse set of areas, particularly in optics and photonics. The intriguing physical effects enabled by non-Hermitian physics and PT symmetry have enhanced significant applications prospects and engineering of novel materials. In addition, it has observed increasing research interests in many emerging directions beyond optics and photonics. This Review paper attempts to bring together the state of the art developments in the field of complex non-Hermitian physics based on PT symmetry in various physical settings along with elucidating key concepts and background and a detailed perspective on new emerging directions. It can be anticipated that this trendy field of interest can be indispensable in providing new perspectives in maneuvering the flow of light in the diverse physical platforms in optics, photonics, condensed matter, opto-electronics and beyond, and offer distinctive applications prospects in novel functional materials.

# 1. Introduction

One of the major research challenges in optics and photonics is to achieve novel and intriguing effects and phenomena in functional platforms with enhanced controllability at will. New approaches that can provide more design freedoms are therefore highly solicited. Thanks to the great advancements of lithography and other nanofabrication procedures during the last three decades, building nanostructures has become realistic which has pioneered substantial progress in artificial device engineering of optical and photonic structures. By judicious selection of materials and designing of the structures, researchers now can achieve all four quadrants of electromagnetic responses: $\varepsilon > 0, \mu > 0$; $\varepsilon < 0, \mu > 0$; $\varepsilon > 0, \mu < 0$; $\varepsilon < 0, \mu < 0$, where $\varepsilon$ is the electric permittivity, and $\mu$ is the magnetic permeability of the medium (**Figure 1**a). [1] Many exotic electromagnetic responses, including negative refractive index ($n<0$), [2-6] negative magnetic permeability ($\mu < 0$), [7,8] and zero refractive index ($n \approx 0$) have been demonstrated. [9-11] However, prior to the pioneering works of Christodoulides *et al.*, [14,15,19,42] most studies only dealt with the real part of the refractive index which, of course has led to a significant number of remarkable achievements in photonics, such as photonic crystals, [16,86,87] periodic media, [20,48] and metamaterials. [21,88] In reality, the refractive index of materials is more than the real part. As shown in **Figure 1**b, if we only consider the electric permittivity $\varepsilon$, there exist four quadrants, in which both the real and imaginary parts of permittivity could be positive or negative. The positive or the negative imaginary parts ($\varepsilon''$) represent the gain or loss of the medium, respectively. In practical design and engineering of an optical system one can resort to the three basic ingredients of nature, namely the refractive index, gain, and loss. While various gain mechanisms have been utilized to enhance weak signals and index contrast to their advantages, loss on the other side is usually considered as unsolicited element since it degrades the efficiency of the optical systems. It is probably for this particular reason that optics and photonics community did not consider exploring simultaneous presence of gain and loss as useful ingredients in device and materials engineering. Recently, an alternative viewpoint aiming at manipulating absorption through a judicious design which involves delicate balance of amplification and absorption processes is receiving considerable research attention. Here one seeks to manipulate the complex refractive index, utilize loss to control gain mechanisms and optical properties for new physical effects and provide new design freedoms in device engineering. Recent explorations of non-Hermitian physics based on parity-time (PT) symmetry offer a novel approach to advance optics toward this goal. While symmetry principles dictate conservation laws and physical behaviors of the systems in nature, intriguing set of physical effects and phenomena emanate from spontaneous symmetry breaking. [12] In regard to PT symmetry as we will see in the course of this Review, this has given rise to achieving new classes of synthetic materials and structures with new physical behaviors and novel functionalities which are seldom observed in traditional structures and which have been experimentally demonstrated, including optical phase transition, [13-19] band merging effects, [14,16] unidirectional light transport, [20-22] PT coherent perfect absorber (CPA), [23-24] and PT CPA-laser. [25-28]

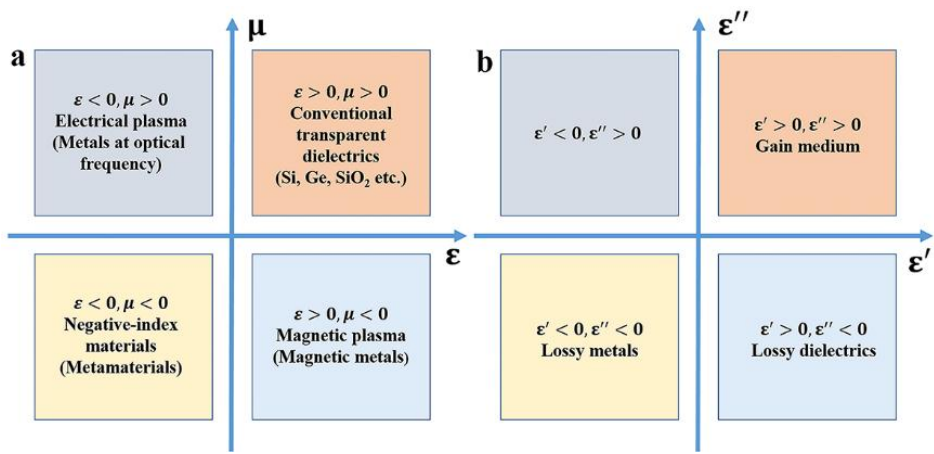

**Figure 1. Overview of different materials.** a) Electric permittivity ($\varepsilon$) and magnetic permeability ($\mu$) form four quadrants that represent the entire range of the electromagnetic response. All four quadrants can be covered by specifically designed materials *without* taking the imaginary part into consideration. b) Complex space of electric permittivity ($\varepsilon$). The positive/negative sign of imaginary part of permittivity ($\varepsilon''$) represents gain/loss in the materials.

The Review of the state of the art developments in the newly emerging field of non-Hermitian physics and PT symmetry has been presented in this paper, in a wide range of areas, predominantly in optics and photonics. We briefly review the basic concepts of PT symmetric Hamiltonians in quantum mechanics and its subsequent extension to optical systems. Next, the basics of PT symmetry, including the parity and time-reversal operations, exceptional points (EPs) and phase transition along with some key effects reminiscent of PT symmetry, for example, asymmetric behaviors, coherent perfect absorption (CPA) and laser, nonlinear phenomena, have been discussed. The recent key developments achieved in a wide range of PT symmetric systems (predominantly optical and photonic platforms) are presented in detail, such as waveguides, microresonators, photonic lattices and crystals, metamaterials, plasmonic systems, hybrid photonic and other systems (see **Figure 2**). We discuss how non-Hermitian physics based on PT symmetry has enabled intriguing physical effects with significant applications prospects for novel materials. Another key feature of this Review is to provide elaborate discussion on new emerging directions of research. Perspective and outlook based on the recent developments have been elucidated along with future directions of research in many new emerging fields. Finally, we conclude.

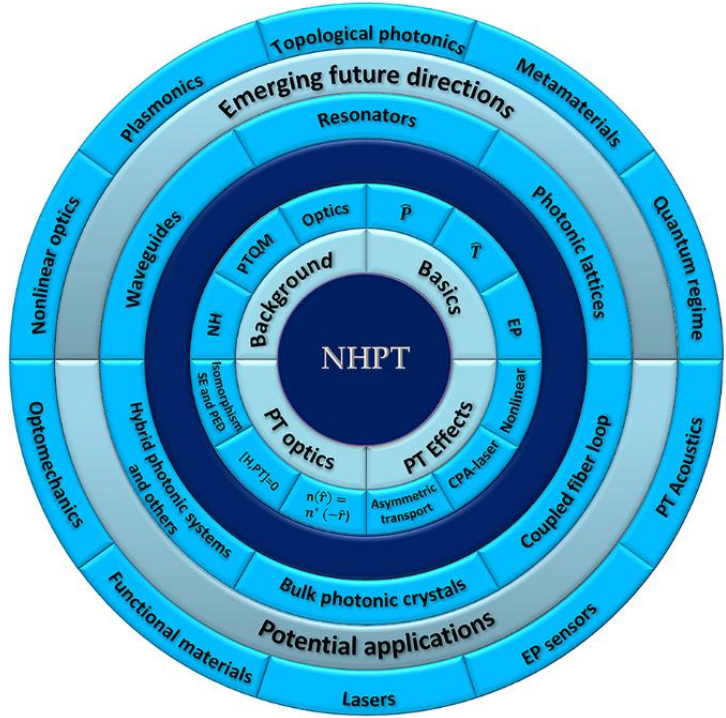

**Figure 2:** An overview of the Review based on non-Hermitian physics and PT symmetry (NHPT).

## 2. Basics of PT symmetry

### 2.1 Basic concepts

PT symmetry was initially proposed in quantum mechanics (PTQM) by Bender and Boettcher in 1998 as an alternative criterion for non-Hermitian Hamiltonians $\hat{H}^\dagger \neq \hat{H}$ († denotes the Hermitian conjugate) that possess a real spectrum. [29-31] They showed that it is not necessary for every operator to obey the Dirac Hermiticity in order to have a physical observable. It was argued that quantum systems with non-Hermitian Hamiltonians can have a real spectrum if they possess PT symmetry. In quantum mechanics, the action of the parity operator $\hat{P}$ amounts to $\hat{P}$: $\hat{r} \to -\hat{r}, \hat{p} \to -\hat{p}$, and that of the time-reversal operator $\hat{T}$: $\hat{p} \to -\hat{p}, i \to -i, \hat{r} \to \hat{r}$, where $\hat{r}$ and $\hat{p}$ stand for the position and momentum operators, respectively. In fact, PT symmetry emerges as a discrete symmetry when the Lorentz group is extended into complex plane. [32] In effect, PT symmetry of the Hamiltonian $\widehat{H}$ implies the following three criteria: first, it commutes with $\widehat{P}\widehat{T}$ as $[\widehat{H}, \widehat{P}\widehat{T}] = 0$, thus $\widehat{H}$ and $\widehat{P}\widehat{T}$ may share a common set of eigenfunctions; second, exact PT unbroken phase and third, self-adjoint Hamiltonian $\widehat{H}$ with respect to the PT inner product. Moreover, the unbroken PT phase points toward the existence of a Hermitian Hamiltonian $\widehat{H}_{Her}$ which possesses the same real eigenspectrum as the non-Hermitian PT symmetric Hamiltonian $\widehat{H}_{PT}$ via an everywhere-defined, invertible and Hermitian operator known as the pseudo-Hermiticity operator $\hat{\eta}$ satisfying $\hat{\eta}\widehat{H}_{PT} = \widehat{H}_{PT}^{\dagger}\hat{\eta}$. [33,34] The real eigenspectrum of the PT symmetric Hamiltonian $\widehat{H}_{PT}$ refers to the positivity of $\hat{\eta}$ and the Hermitian Hamiltonian can be calculated by $\widehat{H}_{Her} = \hat{\eta}^{1/2}\widehat{H}_{PT}\hat{\eta}^{-1/2}$ where $\hat{\eta} = \sum_n |\varphi_n><\varphi_n|$ in a finite dimensional Hilbert space spanned by the state vector $|\varphi_n>$. In fact, non-Hermitian Hamiltonians with real eigenspectra encompass a wide range of Hamiltonians called crypto-Hermitian for which the non-Hermitian Hamiltonian can be projected into a Hermitian one via a similarity transformation. [35,36] We note that the studies on the reality of the eigen-spectra of the non-Hermitian Hamiltonians date back to some earlier time, [37] including early investigations on the Reggeon field theory, [38] and a work by Caliceti *et al.* [39] In their seminal paper, [31] Bender and Boettcher showed via perturbative delta expansion method that the real and positive eigen-spectra of the above-mentioned Hamiltonians were due to the PT symmetry. [40] In a class of quantum-mechanical Hamiltonians $H = \hat{p}^2 + m^2\hat{x}^2 - (i\hat{x})^N$, where $N$ is real, $m$ is mass, various phases of the system were discussed with respect to $N$ and $m$. When $m = 0$, the spectrum of $H$ shows three different spectral behaviors: i) for $N \geq 2$, $H$ results into an infinite, discrete, entirely real and positive eigen-spectrum, ii) for $1 < N < 2$ this produces finite number of real and positive eigenvalues while an infinite number of complex conjugate pairs of eigenvalues, iii) for $N \leq 1$ it gives entirely complex eigen-spectra. $N = 2$ denotes the phase transition point. For $m \neq 0$, the spectrum of H involves another transition at $N = 1$ in addition to $N = 2$.

For a Hamiltonian $\hat{H} = \frac{\hat{p}^2}{2m} + V(\hat{r})$, where m is the mass and V is the potential, to be PT symmetric, a necessary (but not sufficient) condition is $V(\hat{r}) = V^*(-\hat{r})$ (which results from $[\hat{H}, \hat{P}\hat{T}] = 0$), indicating that the real part of the potential is an even function of the coordinate ($V_R(\hat{r}) = V_R(-\hat{r})$) while the imaginary part is an odd one ($V_I(\hat{r}) = -V_I(-\hat{r})$). This condition further allows us to write the Hamiltonian as $\hat{H} = \frac{\hat{p}^2}{2m} + V_R(\hat{r}) + i\mu V_I(\hat{r})$, where $V_{R,I}$ are the symmetric and antisymmetric components of V, respectively. [14,41,42] When $\mu = 0$, the Hamiltonian is Hermitian and the eigenvalues are real. As $\mu$ varies, the eigenvalues may change from real to complex. For $\mu < \mu_{th}$, even if the antisymmetric imaginary part is finite, the spectrum is still real. For $\mu > \mu_{th}$, the spectrum is not real anymore and starts to involve imaginary eigenvalues. So, there exists a critical point $\mu = \mu_{th}$ signifying the onset of spontaneous PT symmetry breaking, which is called the exceptional point (EP). [13,31,43,44] Notably, it is also the point where the eigenvectors of the operator $\hat{H}$ cease to retain completeness as opposed to the Hermitian case (where only eigenvalues coalesce at the so-called diabolical point DP). At this non-Hermitian degenerate point, two or more eigenvalues and their corresponding eigenvectors coalesce and the dimensionality collapse occurs in the eigenspace.

It is worthwhile to note that in much of the existing literature on PT symmetric theory the time-reversal has been implicitly assumed to be even (i.e. $\hat{T}^2 = 1$ for Bosonic systems), whereas it can also be odd for Fermionic systems where $\hat{T}^2 = -1$. [45] It can easily follow from the fact that $\hat{T}$ is an anti-linear operator. Therefore, for a linear operator $\hat{L}$ one gets $\hat{T}\psi = \hat{L}\psi^*$ and $\hat{L}\hat{L}^* = \exp(i\theta)$, assuming that $\hat{T}^2$ only introduces a phase factor in the states ($i.e.\ \hat{T}^2\psi = \exp(i\theta)\psi$), where $\psi$ is a complex component column vector. This results into $\exp(i\theta) = \pm 1$ that corresponds to even and odd time-reversal operations. [45] For example, the even time-reversal operation for a plane wave $\psi = A \exp(iwt - ikr)$ is represented by $\hat{T}\psi = \psi^*(r, -t) = A^* \exp(iwt + ikr)$, which refers to an oppositely moving wave. Parity, on the other hand, is a linear operator $\hat{P}\psi = S\psi$ that commutes with the time-reversal operator, $[\hat{P}, \hat{T}] = 0$ where $S^2 = 1$ and its eigenvalues are $\pm 1$. The inner product for the PT systems reads as $(\varphi, \psi)_{\hat{P},\hat{T}} = (\hat{P}\hat{T}\varphi)^T \psi = \varphi^\dagger S \psi$ without any implicit assumption of orthonormality giving a real but sign indefinite norm $(\varphi, \varphi)_{\hat{P},\hat{T}}$ making it not useful for quantum mechanics. The unitary time evolution of the states in non-Hermitian PT symmetric systems was attributed to the hidden symmetry represented by the linear $\hat{C}$ operator ($\hat{C}\psi = D\psi$, $[\hat{C}, \hat{P}\hat{T}] = 0$) in the unbroken PT phase which forms the CPT inner product (($\varphi, \psi)_{\hat{C}\hat{P}\hat{T}} = (\hat{C}\hat{P}\hat{T}\varphi)^T \psi = \varphi^\dagger D^T S \psi$, in place of Dirac inner product) to ensure unitarity.

## 2.2 PT symmetry in optics

Due to the isomorphic equivalence between the Schrödinger equation in quantum mechanics and the paraxial wave equation, optics has become an ideal platform for exploring PT symmetry. [13,14,41,42,46] Comparing paraxial equation of diffraction (PED) and Schrödinger equation (SE) (Equations (1) and (2)) and by replacing the $t$, $\hbar$ and $V(\vec{r})$ with the $z$, $1/k_0$, and $\Delta n$, we can easily obtain the condition of PT symmetry in optics:

$$i\frac{\partial E(x,z)}{\partial z} = -\frac{1}{2nk_0}\frac{\partial^2 E(x,z)}{\partial x^2} + k_0 \Delta n E(x,z) \qquad (1)$$

$$i\hbar\frac{\partial \Psi(\vec{r},t)}{\partial t} = \left[-\frac{\hbar^2}{2m}\nabla^2 + V(\vec{r})\right]\Psi(\vec{r},t) \qquad (2)$$

where $k_0 = 2\pi/\lambda$, $\lambda$ and $n$ are the wavelength of light in vacuum and refractive index, respectively. Importantly, $\Delta n = n_R(x) + in_I(x)$ is the complex refractive index modulation that plays the role of a complex optical potential. Therefore, a PT symmetric optical system can be achieved by tailoring the complex refractive index distributions satisfying the conditions of PT symmetry $n_R(x) = n_R(-x)$, and $n_I(x) = -n_I(-x)$ (**Figure 3**b). It should be noted that, in optics, the real part of the refractive index $n_R(x)$ signifies dispersion while the imaginary part of refractive index $n_I(x)$ indicates gain or loss. [47] In fact, judicious engineering of the index profiles necessitates precise design as well as fine tuning of the system parameters to establish PT symmetry.

While it is difficult to directly implement experiments on PT symmetry in quantum mechanics, introducing PT symmetry into the classical optics system provides a feasible way. [14,19,48] A unique PT phase transition has been demonstrated in several optical configurations, [15,19] enabling peculiar optical effects such as asymmetric light transport, [16,20,21,48-50] and coherent perfect laser absorber, [25-27] single-mode lasing, [51,52] loss-induced suppression and revival of lasing response, [53] and so on.

## 2.3 PT phase transition and Exceptional Points (EPs)

The abrupt phase transition due to spontaneous breakdown of PT symmetry is an intriguing characteristic of PT symmetric systems. Klaiman *et al.* and Makris *et al.* theoretically investigated PT symmetric optical systems, and found that there exists an abrupt phase transition point, above which dramatically different optical behaviors can be observed. [13,14]

To elucidate the fundamental property of this PT phase transition, we choose the simplest case of a PT directional coupler, as an example. This is schematically shown in **Figure 3**. Using coupled mode theory, the optical fields in the system can be described as:

$$i\frac{d}{dz}\begin{pmatrix}a\\b\end{pmatrix} = \begin{pmatrix}-ig & \kappa\\ \kappa & ig\end{pmatrix}\begin{pmatrix}a\\b\end{pmatrix}. \tag{3}$$

In Equation (3) '$a$' and '$b$' are the slowly varying amplitudes of the optical fields in the first waveguide with loss and second waveguide with gain, $g$ and $k$ are the normalized gain/loss and linear coupling parameters, $z$ is the propagation distance without scaling. The Hamiltonian and the parity operator of the system are given by:

$$H = \begin{pmatrix}-ig & k\\ k & ig\end{pmatrix}, P = \begin{pmatrix}0 & 1\\ 1 & 0\end{pmatrix}. \tag{4}$$

Direct diagonalization of the above Hamiltonian yields the following eigenvalues of the system that show square-root branching behavior:

$$\lambda_\pm = \pm\sqrt{\kappa^2 - g^2}. \tag{5}$$

From Equation (5), it is easily seen that for $g < \kappa$ the eigenvalues are real corresponding to PT unbroken phase, while for $g > \kappa$ the eigenvalues are imaginary, corresponding to PT broken phase. In the unbroken PT phase, the supermodes can be calculated as $|1,2> = (1, \mp exp(\pm i\theta))$ where $\theta = sin^{-1}(g/k)$, whereas in the broken PT-phase they are $|1,2> = (1, -i\ exp(\pm\theta))$ with $\theta = cosh^{-1}(g/k)$. Here $g \equiv g_{th} = \kappa$ refers to the critical point at which both the eigenvalues become zero with the eigenmodes coalescing into $|1,2> = (1, -i\ )$ and becoming self-orthogonal. [29] This indicates a phase transition from PT unbroken phase to broken phase and a second order exceptional point (see **Figure 3**d).

A rather intuitive picture of the two-mode PT symmetric model could be found in the source-sink model. [54] Let us consider two boxes positioned at $x = -a$ and $x = a$ (which in reality refer to the waveguides). The source at a position $x = -a$ radiates energy at the same rate as the absorption of the sink at its mirror image position at $x = a$ (see **Figure 3**c). So, these two boxes denote gain (radiation) and loss (absorption) processes of the whole system. It can also be seen easily that the system is PT symmetric under the joint operations of parity and time-reversal operations. It is because under the parity operation, the gain and loss boxes switch their positions w.r.t. the symmetry axis, and under the time-reversal operation, the gain and loss processes interchange. If the system is isolated i.e. the boxes are not coupled sufficiently strong enough to each other (for which in the two-mode PT dimer model we would have $k < g$), then the system cannot be in equilibrium. It is because the energy in the left box decays down to zero whereas the energy in the right box grows to infinity. The corresponding energies in the two modes are complex (broken PT phase). Now, if the two boxes are coupled strongly enough (for which one can have $k > g$), then the system can equilibrate and the corresponding energy will be real, which oscillate stably between the two waveguides (unbroken PT phase).

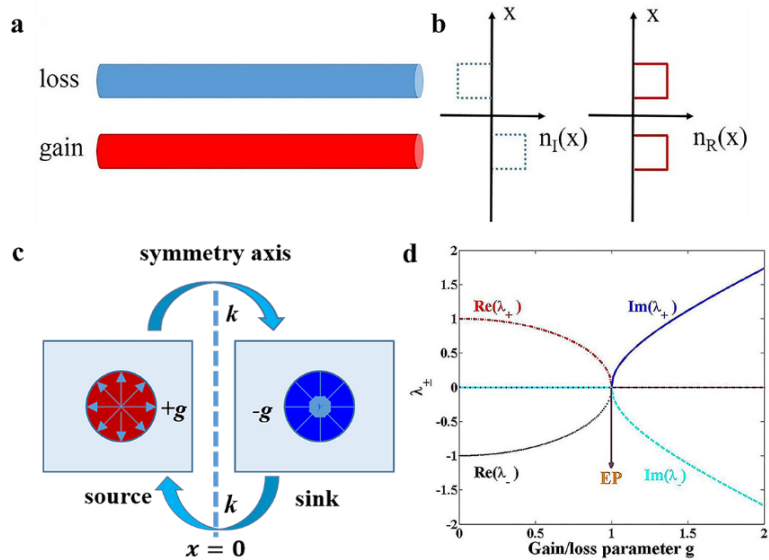

**Figure 3. Basic understanding of a PT symmetric system. a)** Schematic diagram of a two-mode PT symmetric directional coupler. **(b)** The real (imaginary) part of refractive index is an even (odd) function of position. **c)** The simple demonstration of a gain (source)-loss (sink) two

site PT symmetric system, as discussed in the ref. [54] **d)** Different branches of the eigenspectra of the PT symmetric dimer with the coupling constant $k = 1.0$ in Equation (5), showing the EP as the point of non-Hermitian degeneracy.

It is worth mentioning that exceptional points, as branch point degeneracies are essential not only in fundamental understanding of the properties reminiscent of generalized set of non-Hermitian systems, but also in introducing many intriguing effects. These include enhanced mode-splitting for ultrasensitive sensors via EPs, [55-57,101] chiral mode conversion, [69,70,58] mode discrimination in multimode lasers, [52] single-mode lasing, [51,52] unidirectional invisibility, [16,20,21] loss-induced suppression and revival of lasing, [53,98] and many more. Although, the physics of exceptional points can be observed and exploited in a large number of physical settings, optical and photonic integrated structures provide one of the versatile platforms. [59] Some works investigate reconfigurable PT phase transition via time-dependent periodic Floquet driving that makes it possible to achieve spontaneous PT breaking at arbitrary value of the non-Hermitian parameter. [60,61]

A number of theoretical propositions were put forward in exploring the possibility of realizing PT symmetry in practical optical settings, [41,62-64] in particular the work of El-Ganainy *et al*. [41] firmly established the theoretical framework for a PT symmetric optical coupler and paved the way for experimental realizations of PT symmetry in real physical settings. Afterward, Guo *et al*. first experimentally observed PT symmetry breaking in lossy coupled optical waveguides system and a loss-induced optical transparency beyond the phase transition point was realized. [19] Experimental demonstration of balanced gain and loss PT symmetry was reported by Rüter *et al*. in coupled optical waveguides. [15] Practically speaking, realization of an ideal gain-loss balanced PT system could be difficult not only because of the limited gain bandwidth but also due to ineluctable manufacturing and fabrication imperfections. For instance, the physical system of a coupled imbalanced lossy resonators constituting a metamolecule of coupled Lorentzian dipoles $P_0 = (\tilde{P}_x, \tilde{P}_y)e^{iwt}$ resonating at $w_0$ to the incident fields of $E_0 = (\tilde{E}_x, \tilde{E}_y)e^{iwt}$ which can equivalently be described via a gauge transformation by a gain-loss PT symmetric system placed on a background lossy medium. Under certain conditions ($\delta \ll w_0, \gamma_y < \gamma_x \ll w_0$) the system can be represented by: [67,88,130]

$$S\begin{pmatrix}\tilde{p}_x\\ \tilde{p}_y\end{pmatrix} + \begin{pmatrix}-i\gamma & g_{xy}\\ g_{xy} & i\gamma\end{pmatrix}\begin{pmatrix}\tilde{p}_x\\ \tilde{p}_y\end{pmatrix} = g\begin{pmatrix}\tilde{E}_x\\ \tilde{E}_y\end{pmatrix} \qquad (6)$$

where $S = \delta + g_{xx} + i\gamma_{avg}, \gamma_{avg} = (\gamma_x + \gamma_y)/2$ is the average loss, $\gamma = (\gamma_y - \gamma_x)/2$ is the loss contrast, $g_{xy}$ and $g_{yx}$ are contributions from retarded field coupling, $g$ is the strength of incident fields, and $\delta = w - w_0$. In this way the eigenstates structure is solely determined by the PT symmetric term (second term on the left-hand side of Equation (6)) in the equation and the first term only imposes a damping background, in other words, an ideal PT symmetric system embedded in a lossy medium. As anticipated, varying $g_{xy}$ different PT phases were identified. This type of passive PT symmetry could be particularly relevant in efficient and controllable realization of PT related physics and phenomena in practical settings where incorporating balanced gain-loss framework is challenging.

Besides these works, PT phase transition has also been observed in many other physical systems and scenarios, such as asymmetric diffraction at the exceptional point, [65,66] coherent perfect absorber, [67,23-28] optical isolation, [22,68] and so on.

## 2.4 Asymmetric light transport

Asymmetric transport of light waves poses a vast range of practical applications in effective engineering of many optical and photonic devices. Especially, on-chip and integrated optical components which provide asymmetric response, find immense potential research attention. It turns out that optical and photonic systems can be engineered to this aim via the notions of non-Hermitian physics and PT symmetry. At the phase transition point (EP), the eigenmodes in the PT symmetric periodic structures become degenerate, leading to asymmetric light transport. [16,20-22,48,68] Lin *et al.* proposed to combine optical gain and loss modulations together with refractive index perturbations to form a PT-synthetic Bragg structure, shown in **Figure 4**a, and have theoretically demonstrated unidirectional reflection at the EP. [20] The unidirectional invisibility stems from the spontaneous PT symmetry breaking at the EP, as opposed to the conventional cloaking devices in which one considers scattering of waves from a cloak medium.

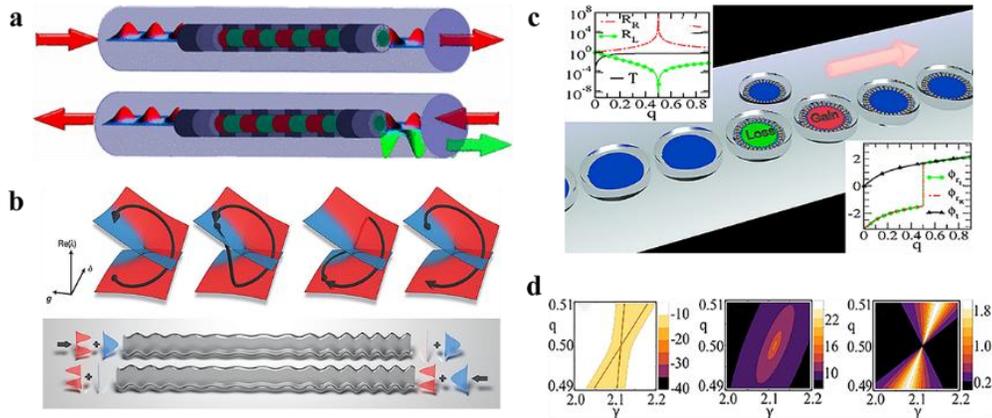

**Figure 4. Asymmetric transport of light. a)** Schematic illustration of unidirectional reflection in a PT symmetric Bragg scattering device. [20] **b)** Trajectories corresponding to the different encircling directions for the two Riemann sheets (red and blue surfaces) have been shown to show the non-adiabatic nature while dynamically encircling an EP. [70] **c)** Schematic representation of the coupled PT symmetric disk resonators (upper panel). [72] The insets show the reflections, transmission coefficient, and the phases of the transmission and reflection amplitudes versus the wave vector q. **d)** Density plots of the logarithm of the left and right reflection, and the transmission coefficient. [72] Figures reproduced with permissions from: a) ref. [20] Copyright 2011, American Physical Society; b) ref. [70] Copyright 2016, Springer Nature; c,d) ref. [72] Copyright 2014, American Physical Society.

For a simple demonstration of asymmetric behavior due to PT symmetry, we consider a PT symmetric grating with the refractive index distribution $n(x) = n_0 + n_1 \cos(2\beta x) + in_2 \sin(2\beta x)$ when $|x| < L/2$ (background refractive index $n_0$ for $|x| > L/2$ with $n_1 \ll n_0$, $n_2 \ll n_0$ and $\beta$ is the grating wavenumber $\beta = \pi/\Lambda$, $\Lambda$ is the grating period), for which the electric field $E(x,t) = E(x)\exp(-j\omega t)$ ($\omega$ is the angular frequency), satisfies the Helmholtz equation:

$$\frac{d^2 E(x)}{dx^2} + \left(\frac{\omega}{c_0}\right)^2 \epsilon(x) E(x) = 0 \tag{7}$$

where $c_0$ is the speed of light in vacuum. It is found that the transmission is the same for left or right incidence, while the reflection may differ. In Hermitian case when $n_2 = 0$, the reflection is the same, $R_L = R_R$, for left or right incidence. At the Bragg point of the device $\delta = 0$ ($\delta = \beta - k$), the transmission goes to zero while the reflection becomes unity in the large $L$ limit. When $n_2 \neq 0$, asymmetric reflection starts to develop. The most remarkable case occurs when $n_1 = n_2$, $\delta = 0$, where the reflection and transmission from the left incident wave are zero and unity, respectively. At the same time, the reflection from the right incident wave follows quadratic relation with the sample length boosting reflection at the exceptional point: $R_R \xrightarrow{\delta \to 0} L^2 \left(\frac{kn_1}{n_0}\right)^2$.[20] Such a phenomenon is referred to as unidirectional reflectivity (**Figure 4**a). This effect was shown to be valid over all frequencies around the Bragg point, even in presence of nonlinearity. It is worthwhile to note that asymmetric reflection does not indicate nonreciprocal light transport. Moreover, asymmetric reflection does not point to asymmetric transmission or nonreciprocity. For instance, in regard to non-Hermitian systems, in a recent work,[69] to show chiral emission and directional lasing at an EP it was found that even though it depicts asymmetric reflection but the transmission is bidirectional, hence reciprocal. In linear, time-independent and nonmagnetic systems, PT symmetry or PT breaking cannot induce nonreciprocal effect. This is due to the fact that gain and/or loss elements in non-Hermitian systems breaks the time-reversal symmetry but not reciprocity for which bidirectional transmission and asymmetric reflection follow. On the other hand, when Lorentz reciprocity is broken due to some other means (for example, in presence of nonlinearity), it results into asymmetric transmission and hence, nonreciprocity.

The asymmetric light transport has witnessed many experimental demonstrations, [16,21,22,68] including some interesting optical settings (**Figure 4**). [20,70-73] Ramezani *et al.* proposed a two-mode Kerr-nonlinear coupled waveguides system to study directed wave transport. [77] It was found that due to interplay between PT symmetry and nonlinearity there is a critical value of the nonlinearity strength beyond which unidirectional behavior occurs. Some experimental works demonstrate nonreciprocal response in systems of coupled resonant microcavities via judicious interplay of PT symmetry and nonlinearity where PT breaking results into enhanced gain saturation nonlinear effect and nonreciprocity. [22,68] In such cases, the nonreciprocity stems from enhanced gain-saturation nonlinearity of the active resonator due to spontaneous PT breaking. So, PT symmetry-assisted effects (PT breaking) may provide useful means toward achieving nonreciprocity in such synthetic resonant structures (see Figure 10). On the other hand, we note that non-Hermitian elements present in a non-Hermitian system result in non-adiabatic transitions, which in turn give rise to chiral behaviors. However, observing this effect experimentally witnessed some key issues due to the requirement of fully dynamical encircling. A waveguides system with two transverse modes has been demonstrated to show the above-mentioned idea of chiral mode transport. [70] Whereas, in adiabatic case, the state-flip is observed upon encircling an EP, the breakdown of adiabaticity in non-Hermitian systems affects the directionality of EP encirclement, and the state-flip is not observed (**Figure 4**b). Another work shows simultaneous unidirectional lasing and unidirectional reflectionless behaviors in a one dimensional chain of coupled microcavities with a pair of PT symmetric defects at the center. [72] Diverging reflection from the gain disk (pink arrow) and the almost zero reflection from the loss disk are found. In addition, at the spectral singularity point, denoted by $(q, \gamma) = (2.1, 0.5)$, while the left and right reflection coefficient tend to zero and infinity respectively, the transmission is a multivalued function (**Figure 4**c). In this connection, it is worthwhile to note that some more general non-Hermitian methods have been proposed to achieve unidirectional reflectionless behavior based on spatial Kramers-Kronig (K-K) relations, [74] and Bragg scattering via judicious index and gain tailoring. [75] In addition to the above demonstrations, asymmetric light transport phenomena have also been reported by several groups, [48,68, 76 - 78] with different optical configurations such as optical couplers, [76,77, 79] two-mode waveguides, [48] and coupled resonators. [68]

## 2.5 Coherent perfect absorption (CPA)-Lasing

Chong *et al*. proposed CPA as the time-reversed counterpart of laser for which the time-reversed coherent and monochromatic radiation of the laser output incident on the resonator filled with lossy medium gets perfectly absorbed. [23] On a more fundamental level it occurs due to combined effect of interference and dissipative effects which account for the radiation trapping and absorption due to dissipation. For example, one may refer to the scalar wave equation $(\nabla^2 + k^2 n(r)^2)\varphi_k(r) = 0$ with $n(r) = n_1(r) + i\, n_2(r)$, $k = w/c$. Here $n_2(r) > 0$ and $n_2(r) < 0$ refer to the absorbing and amplifying cases respectively. The steady-state solutions of the equation could be described in the scattering matrix formalism which for incoming and outgoing wave amplitudes $\alpha$ and $\beta$ gives $S[n(r)k].\alpha = \beta$ and its time-reversed counterpart $S[n^*(r)k^*].\beta^* = \alpha^*$, which essentially means that for every scattering solution in the amplifying case corresponding to the outgoing wave $\beta$, there exists solution for absorbing the case corresponding to the outgoing solution $\beta^*$. [23] This novel effect has further widened the possibility of judicious control and manipulation of absorption in the materials systems.

Shortly after CPA was proposed, Longhi showed that if the optical medium is designed such that it satisfies the PT symmetry condition (say, $\varepsilon(r) = \varepsilon(-r)^*$) for the dielectric function, it can simultaneously act as laser and CPA with appropriate amplitudes and phases at the lasing threshold. [25] While the feature of a laser is its intrinsic coherence both in space and time, through time-reversing its counterpart, CPA, could be generated from a lossy resonator. The complex PT symmetric modulation of the dielectric function is applied in the regions $|x| < L/2$ where the laser oscillator resides. Outside of this region, the dielectric function is assumed to be real (**Figure 5**a).

The coexisting coherent perfect absorption and lasing phenomena in PT symmetric systems that can be observed around the phase transition point (EP), show intriguing possibility enabled by PT related ideas. [24-28] Both laser and CPA constitute important key concepts in optical and photonic systems. However, coexistence of laser and CPA in one cavity is counterintuitive and thus unique in PT symmetric systems, enabling better control over coherent amplification and absorption of light fields.

On the other side, in an InGaAsP multi quantum well waveguides system placed on InP gain substrate with periodic arrangement of attenuating Cr/Ge structures, an ideal gain-loss modulated PT system was realized by uniform pumping of the waveguides system with the observation of lasing and anti-lasing modes at the same cavity. [27] In this experiment, Wong *et al.* have experimentally demonstrated a CPA-laser using periodical arrangement of alternate gain/loss structure along the light propagation direction via judicious phase offsets of the input signals. In **Figure 5**b, the incoming waves have -π/2 phase offset that leads to constructive interference of the resonant fields and the strong confinement of light in the gain regions for the lasing mode. In **Figure 5**c), the incoming waves have π/2 offset which leads to confinement of light in the loss regions and consequently the absorption of the anti-lasing mode. In fact, the coherent incidence conditions in the form of interferometric phase control over the guided light coming from the both directions and periodic modulation of the lossy components (period is half the wavelength of the light) on the InGaAsP/InP gain medium can selectively excite either the lasing or the anti-lasing mode through confining the field distribution inside the gain or loss region, respectively (**Figure 5**d). The good amplification to absorption ratio requires careful engineering of the system to reach the CPA-laser point. It provides a versatile platform for effective coherent manipulation of light in a single synthetic photonic circuit for a wide range of optical effects, including lasing, coherent absorption and amplification.

Moreover, the scattering properties of PT symmetric optical systems have been discussed using scattering matrix formalism. [28] It shows the PT breaking phase transition behavior and the associated eigenstates in the unbroken and broken phases of PT symmetry. **Figure 5**e shows the PT symmetry breaking phase transition occurring at the critical value of the frequency beyond which the scattering intensity witnesses strong oscillation. The phase diagram depicts the distinct phases and the solutions of the CPA-laser as the discrete points in the upper panel of **Figure 5**f. The lower panel shows how insertion of a different real-index medium into the structure can result in altogether different response, in which the resonances within the real-index cause the trapping time to oscillate. Interestingly, this work reveals that the simultaneous occurrence of the lasing and CPA actions stems from the coincidence of the pole and zero of the scattering matrix which occur at the specific points in the broken PT phase.

In manipulating CPA in real applications with absorption, a silicon laser cavity with two counterpropagating incoming waves was studied in an interferometric setup, as shown in **Figure 5**g. [24] Coherent laser radiation from Ti: sapphire laser source impinges on a beam splitter and gets reflected and transmitted through an experimental arrangement consisting of attenuator and phase controlling device. The attenuator and the phase controlling device help in maintaining appropriate amplitudes and phases of the incident beams. The light from either directions finally reach the silicon cavity in proper phases and amplitudes to give rise to CPA due to destructive interference.

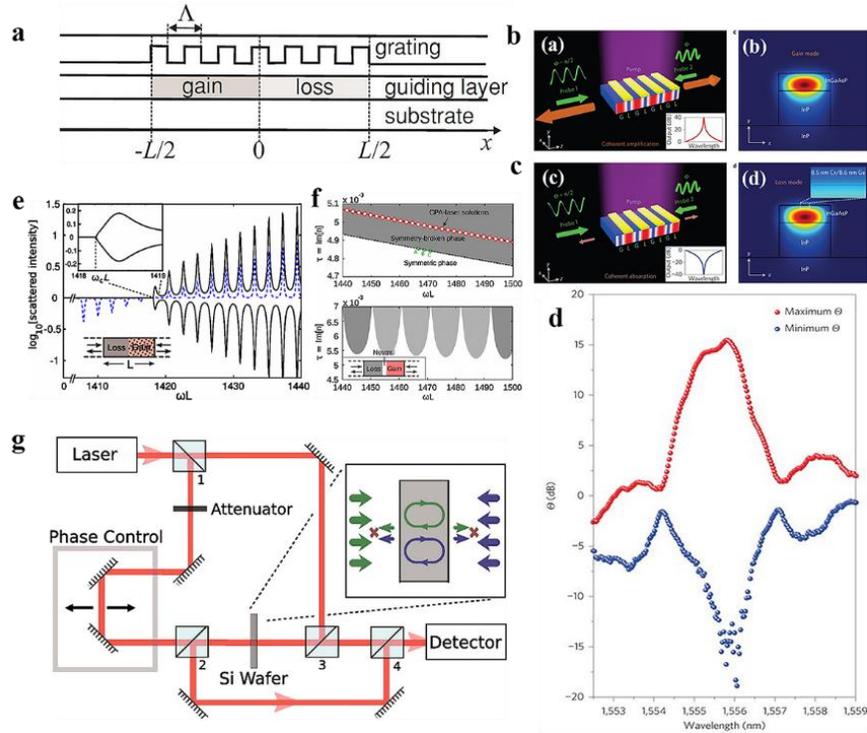

**Figure 5. CPA and lasing. a)** Schematic of a PT CPA laser based on the homogeneous and symmetric gain and loss distributions on a uniform index background. [25] **b,c)** Schematic representation of the gain-loss PT symmetric photonic structures ((a), (c)). [27] Field distributions of the lasing and anti-lasing modes, as shown in (b) and (d). **d)** The plot of the spectra output coefficient of CPA-laser in an InGaAsP multi quantum well waveguides placed on InP gain substrate with periodic arrangement of attenuating Cr/Ge structures.[27] Output spectra coefficient, showing lasing mode (for $-\pi/2$ phase offset, red curve) and anti-lasing mode (for $-\pi/2$ phase offset, blue curve) at the same cavity. **e)** Plot of S-matrix eigenvalue intensities vs the frequency ω (solid curves). Here, total length of the lattice is *L* with gain-loss elements. Inset shows the enlarged view of the PT breaking region. The blue dashed curve denotes minimum scattering intensity for equal incoming waves. **f)** Plot of the phase diagram of the S-matrix eigenvalues for 1D PT symmetric systems. [28] **g)** Schematic diagram of the experimental setup for a silicon laser cavity-based CPA system [24] Figures reproduced with permissions from: a) ref. [25] Copyright 2010, American Physical Society; b,c,d) ref. [27] Copyright 2016, Springer Nature; e,f) ref. [28] Copyright 2011, American Physical Society; g) ref. [24] Copyright 2016, Nature Publishing Group.

## 2.6 Nonlinear PT systems

Apart from gain and loss, another important component of the optical systems is nonlinearity, which gives rise to a plethora of interesting physical effects and behaviors. In fact, nonlinearity and non-Hermitian elements oftentimes coexist in the optical and photonic settings. Degrading dissipative factors are associated with strong nonlinear materials, and so appropriate system designs are solicited in addressing this issue. In particular, the fact that light can self-regulate the index profile through nonlinearity showed possibility of inducing effects on the PT phase transition on different power levels. In PT symmetric systems, when nonlinearity is introduced, the Hamiltonian describing the system does not commute with the PT operator. However a theoretical study showed the existence of stable self-trapped nonlinear stationary soliton solutions despite the presence of gain and loss, [64] as found in several other works. [80] The optical wave propagation in such models is described by the nonlinear Schrödinger equation (NLSE) with a complex PT symmetric periodic potential:

$$i\psi_z + \frac{1}{2}\psi_{xx} + [V_R(x) + iV_I(x)]\psi(x,z) + |\psi(x,z)|^2\psi(x,z) = 0 \qquad (8)$$

$\psi(x,z)$ is proportional to the electric field, $x$ representing the transverse co-ordinate and $z$ the normalized propagation distance. The linear potential is PT symmetric in the usual sense that $V_R(-x) = V_R(x)$ and $V_I(-x) = -V_I(x)$. Equation (8) in its linear case yields the (quasi) power conservation relation as $Q = \int_{-\infty}^{\infty} dx\, \psi(x,z)\psi(-x,z)$ in contrast to the EM power relation $P = \int_{-\infty}^{\infty} dx\, |\psi(x,z)|^2$. It was shown that in the presence of nonlinearity these quantities satisfy the evolutions equations: $i\frac{dQ}{dz} = \int_{-\infty}^{\infty} dx\, \psi(x,z)\,\psi^*(-x,z)\,[|\psi^*(-x,z)|^2 - |\psi(x,z)|^2]$ and $i\frac{dP}{dz} + 2\int_{-\infty}^{\infty} dx\, |\psi(x,z)|^2 w(x) = 0$. At higher power levels, despite the imaginary

component staying above the linear PT threshold, nonlinear modes can continue to exist as it nonlinearly shifts the threshold value and the spectrum is real in contrast to the low power levels where the system remains in the broken phase. Such nonlinear PT symmetric systems can host stable self-trapped solitons, as has been depicted in **Figure 6**a that shows the stable soliton solution in Scarf II potential under appropriate parameter values ($V(x) = V_R + iV_I = V_0 sech^2(x) + W_0 sech(x) \tanh(x)$, for $V_0 = 1$, $W_0 = 0.5$). Lumer *et al.* showed the effects of nonlinearity in inducing or suppressing the PT phase transition in a photonic lattice system.[81] It was predicted that nonlinearity can reverse the PT breaking order from broken PT phase to unbroken one and vice a versa. Linear stability of the nonlinear modes revealed the real eigenvalues suggesting stable nonlinear wave solutions. **Figure 6**b depicts the evolution of the maximal power for all the eigenvalues (solid red) and for the real eigenvalues (blue circles) versus the wave-vector for an initial excitation in presence of nonlinearity. It was found that if the red solid curve is multiplied by 0.5, the two curves coincide. The number 0.5 is not unique and it depends on the PT potential considered. So, this simple demonstration shows how nonlinearity controls the PT phase transition. It was later observed in some other works.[82,83] Nonlinear PT systems have proved to be useful in inducing many important physical effects. It includes, for example, PT phase transition induced by nonlinearity,[81,82,83] optical isolation in coupled microcavities with saturable nonlinearity,[22,68] optical solitons in PT symmetric lattices,[108] and so on. In particular, Wimmer *et al.* demonstrate the possibility of the existence of stable nonlinear localized states or discrete solitons in 2D PT symmetric mesh lattices.[108] The experimental setup considered two coupled fiber loops with slightly different lengths. As shown in **Figure 6**c, under proper choice of the parameter values, the system can give rise to either unstable dispersing wave profile in broken PT phase ((a),(c) in **Figure 6**c) or localized nonlinear modes ((b),(d) in **Figure 6**c) in unbroken PT phase. On the other hand, lasers are an ideal platform to investigate effects enabled by non-Hermitian physics. Moreover, lasers are linear at the first lasing threshold, and become nonlinear beyond it. A recent work studies nonlinear modal interactions in a PT symmetric laser using saturable nonlinearity.[83]

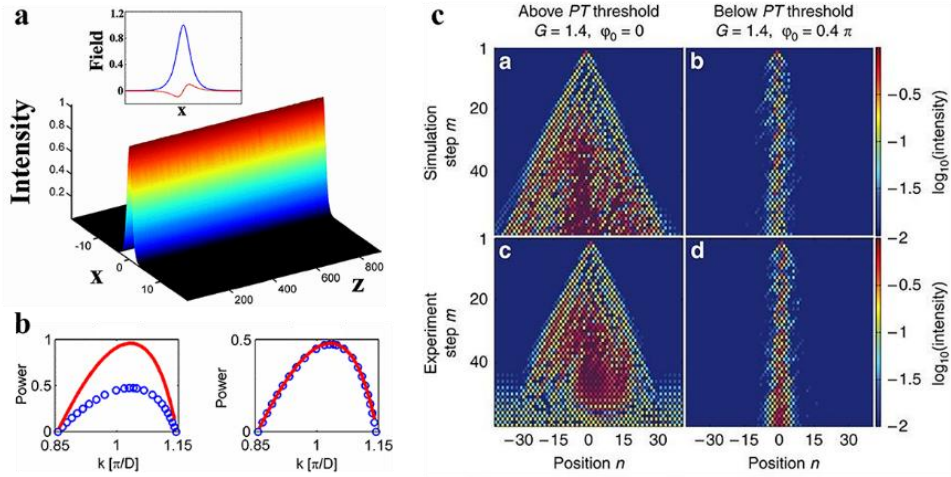

**Figure 6: Nonlinear PT symmetric systems. a)** The stable evolution dynamics of the nonlinear wave solution in a PT symmetric Scarf II potential. Inset: the symmetric (blue solid) and anti-symmetric (red dotted) parts of the fields. [64] **b)** The dynamic power evolution (solid red) and the power evolution corresponding to the real eigenvalues (blue circles) vs wave-vector (left panel). The right panel shows the same as in left panel, [81] but multiplied by 0.5. **c)** The demonstration of the PT soliton formation in a 2D PT mesh lattice structure. [108] (a),(c) Dispersed and unstable wave profile for the initial excitation in broken PT phase. (b),(d) PT soliton profile in the unbroken PT phase. Figures reproduced with permissions from: a) ref. [64] Copyright 2008, American Physical Society; b) ref. [81] Copyright 2013, American Physical Society; c) ref. [108] Copyright 2015, Nature Publishing Group.

A flurry of research interests have emerged in the subfield of nonlinear PT optical systems, for a more comprehensive review see the articles,[84,85] and the references therein.

## 3. Different system configurations

Since the first experimental demonstrations of PT symmetry breaking in optics, [19,15] so far several types of optical structures and configurations have been utilized to construct synthetic optical and photonic PT systems, including waveguides, [13,48] resonators, [22,68] photonic lattices, [16,107-111] photonic crystals, [86,87] metamaterials, [88] plasmonic systems, [89-91] and so on. In what follows, we present key important developments in such optical and photonic platforms and discuss the results and their practical implications. It is likely to provide not only a clear picture of how these different platforms have facilitated to investigate the intriguing physics of non-Hermitian PT symmetric systems, but also demonstrate the advances achieved in their characteristic ways.

## 3.1 Waveguides

Waveguide is one of the basic building blocks in optics. Especially the on-chip manipulation capability attracts a lot of research interests. Due largely to its ability to provide simple and efficient platform to simulate physical processes, the initial investigations related to PT optics started from waveguide systems.

In the first experimental observation of PT symmetry breaking in optics Guo *et al.* considered two coupled optical waveguides built on a multilayer double-well $Al_xGa_{1-x}As$ heterostructure in which chromium along the propagation direction on one waveguide introduced loss (**Figure 7**a). As shown in **Figure 7**b, an expected decay of the total transmission occurred as the loss was increased. However, beyond a critical point where the transmission is nearly zero, the transmission was found to increase as the loss was increased. [19] This anomalous transmission can be understood from the spontaneous PT symmetry breaking at the critical value of the loss parameter. Below this critical point, the system resides in the PT unbroken phase for which the usual decrease in the transmission was observed with a decrease in the loss, followed by increasing transmission beyond the critical point due to PT symmetry breaking.

In the first realization of ideal balanced gain-loss PT symmetry in optics by Rüter *et al.*, [15] based on a Fe-doped LiNbO$_3$ coupled waveguides system (**Figure 7**c), the real part of the refractive index modulation was provided through Ti in-diffusion, the gain through two-wave mixing process and the losses due to the doping elements. **Figure 7**d depicts the inherent PT response of the system. Initially when the first (second) port of the system is excited the corresponding intensity profile, shown in the left (right) panel, exhibits nonreciprocal behavior when gradually gain builds up in the system. Below (above) the PT threshold the intensity shows reciprocal (exponentially growing or decaying) behavior. It should be noted that unlike systems with only losses, systems with balanced gain and loss exemplify ideal PT symmetric systems with real eigenspectra.

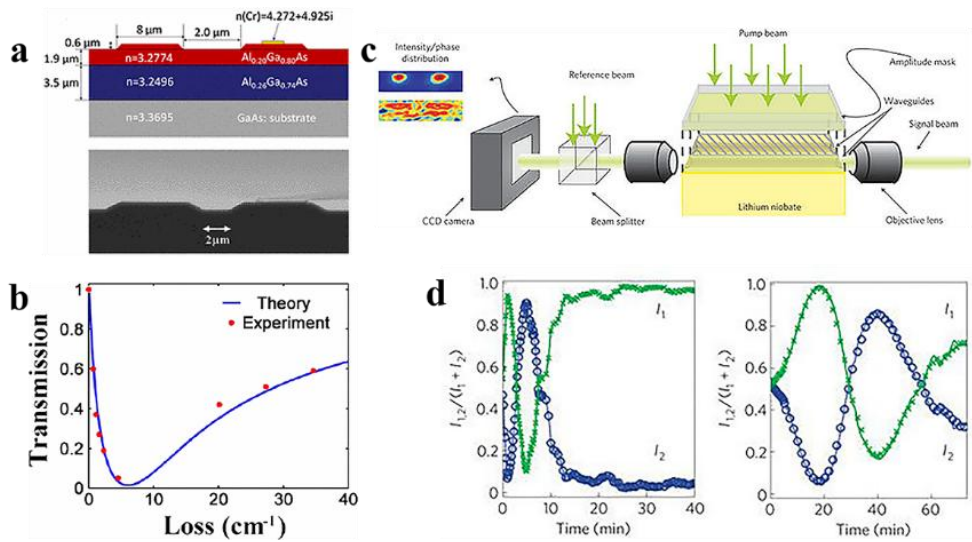

**Figure 7. PT symmetric waveguides: experiments in optics. a)** Practical design of passive PT waveguides (upper panel) and the SEM image (lower panel). [19] **b)** Experimental result of spontaneous passive PT symmetry breaking phenomenon. [19] **c)** Experimental demonstration of the active balanced gain-loss PT symmetric waveguides based on a Fe-doped *LiNbO$_3$* coupled waveguides. **d)** Measurement of the intensities of the PT symmetric coupled waveguides system vs time where the left (right) panel corresponds to initial excitation in the first (second) port [15]. Figures reproduced with permissions from: a,b) ref. [19] Copyright 2009, American Physical Society; c, d) ref. [15] Copyright 2010, Nature Publishing Group.

Effective light by light switching is an important aspect in optical communications and information processing networks. But typically it necessitates high optical power, which hinders it practical applications. The asymmetric reflection near the EP of the non-Hermitian photonic system of a PT metawaveguide (**Figure 8**a) was exploited to obtain low power switching of light in linear regime as opposed to using nonlinearity at the cost of high power requirement. [92] Under appropriate phase modulation of the control, two distinct modes of operations, namely CPA and strongly scattered regime were found (**Figure 8**b) at the resonant wavelength. A weak control beam was interferometrically manipulated to control the behavior of the strong laser light with an extinction ratio 60 dB, providing an efficient way for judicious control and manipulation of light wave guiding, routing and switching. Further consideration of nonlinearity may enable important effects due to broken Lorentz reciprocity.

After theoretical prediction of unidirectional invisibility induced by PT breaking at the EP, [20] unidirectional reflectionless behavior was exhibited in a CMOS compatible passive PT metamaterial embedded on silicon-on-insulator (SOI) platform at optical frequencies, [21] on-chip and in spatial domain. Properly engineered complex dielectric permittivity profile was employed to achieve the PT potentials in this passive metamaterial system. It consists of periodically arranged complex potentials 760 nm in length, placed over 800 nm Si waveguide which is embedded inside $SiO_2$. The real and imaginary parts of the complex potential are realized by 51 nm Si and 14 nm Ge/ 24 nm Cr. The reflectivities for both the forward and backward directions show asymmetric response as is evident from the forward and backward reflectivities in **Figure 8**c and the contrast ratio over a large wavelength range in **Figure 8**d. It is found that the forward reflectivity is about 7.5 dB higher than the backward. **Figure 8**e clearly shows the unidirectional propagation of the light fields only in the forward direction at the EP.

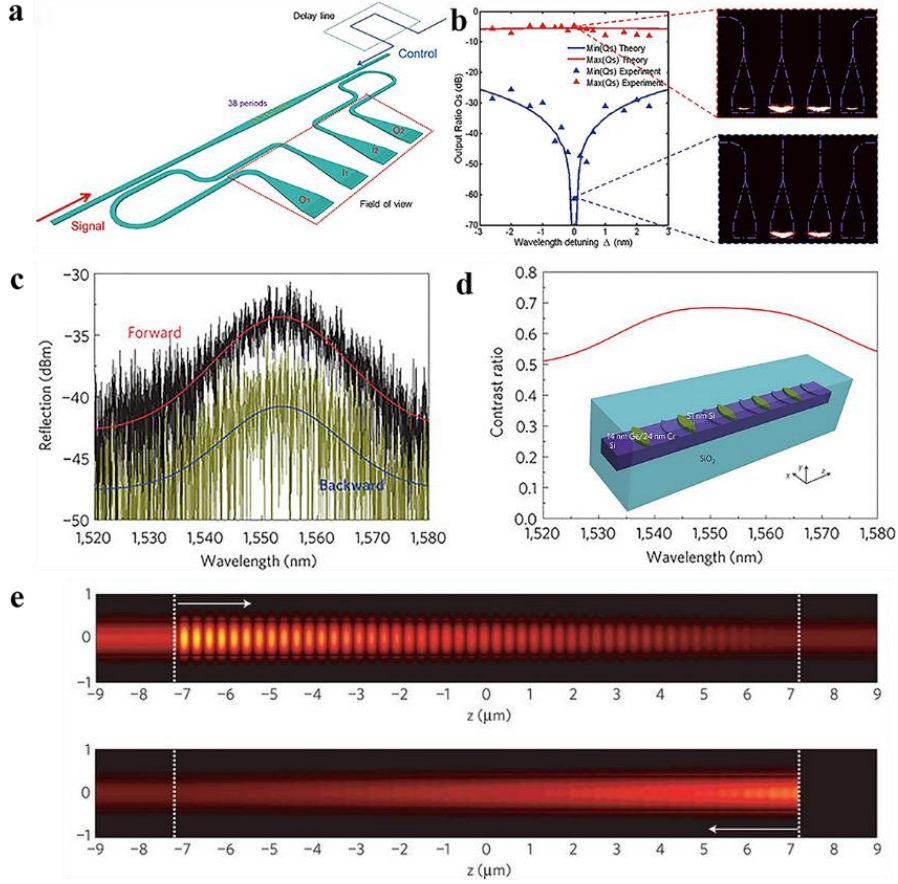

**Figure 8. PT symmetric waveguides: experiments on silicon waveguides. a)** Experimental setup for asymmetric interferometric light-light switching in a PT metawaveguide. **b)** Spectra of maximum (red) and minimum (blue) output scattering coefficients of the system in (a). [92] The right side panels show the intensity snapshots in microscope for strongly scattered (top) and CPA (bottom) at the resonant wavelength. **c)** The measurements of the reflection spectra for forward (black) and backward (green) directions. The red and blue curves represent the Gaussian fits of the raw data. **d)** The contrast ratio of the reflectivities over a large telecommunication range. [21] Here the inset shows the schematic diagram of the passive PT metamaterial system. **e)** The demonstration of the unidirectional propagation of light fields in the PT metamaterials at EP. Light reflects only in the forward direction. [21] Figures reproduced with permissions from: a,b) ref. [92] Copyright 2016, American Association for the Advancement of Science; c,d,e) ref. [21] Copyright 2013, Nature Publishing Group.

In demonstrating one-way asymmetric propagation of light in a passive PT symmetric silicon photonic system, [48] Feng *et al.* consider periodic optical potentials having complex modulations in the dielectric constants $\Delta\varepsilon = \exp(iq(z - z_0))$, $q = k_1 - k_2$, along the propagation direction of light (**Figure 9**a). The system possesses a fundamental symmetric mode ($k_1$) and a higher order anti-symmetric mode ($k_2$). For incoming symmetric mode, the one-way mode conversion could be achieved at the EP for backward wave when the phase-matching condition ($\Delta k = k_2 - k_1 + q$) is satisfied, as shown in **Figure 9**b. In a subsequent study, [95] asymmetric unidirectional propagation of light was experimentally verified in organic thin film waveguides (see **Figure 9**c) via a combination of experimental techniques such as interference lithography and oblique angle deposition, following the observation of passive PT symmetry in silicon waveguides, [21] and in multilayered organic thin films. [93] The diffracted first order light from the left and right directions showed noticeable asymmetry that stems from the passive PT symmetry breaking of the complex refractive index variation $\Delta \tilde{n}(z) = \Delta n \cos(qz) + i\Delta k \sin(qz)$ at the exceptional point $\Delta n = \Delta k$. If properly exploited it could be testbed for exploring PT optics related ideas in large-area active platforms.

Judicious exploitation of the notions pertaining to fundamental physics often prove to be useful in addressing bottlenecks of technological advancements. Harnessing symmetry considerations in practical photonic systems gives rise to novel attributes. For example, supersymmetry (SUSY) from initial domain of particle physics and quantum field theory has spread into many other areas, including optics and condensed matter, and it has promised intriguing possibilities such as robust single-mode lasing. In conventional laser array, multi-mode lasing spectra and nonlinear modal interactions often pose significant challenge. It turns out that one can alleviate this particular issue via strategic utilization of SUSY-related ideas in conjunction with non-Hermitian physics. [94] In such scenarios, the primary laser array (described by a Hamiltonian, say, $H^{(1)}$) is coupled with its dissipative superpartner array ($H^{(2)}$), as shown in **Figure 9**d. Under specific coupling condition, the dissipative superpartner array aids in suppressing all the higher-order modes except the fundamental lasing mode due to which it gets access to the entire gain.

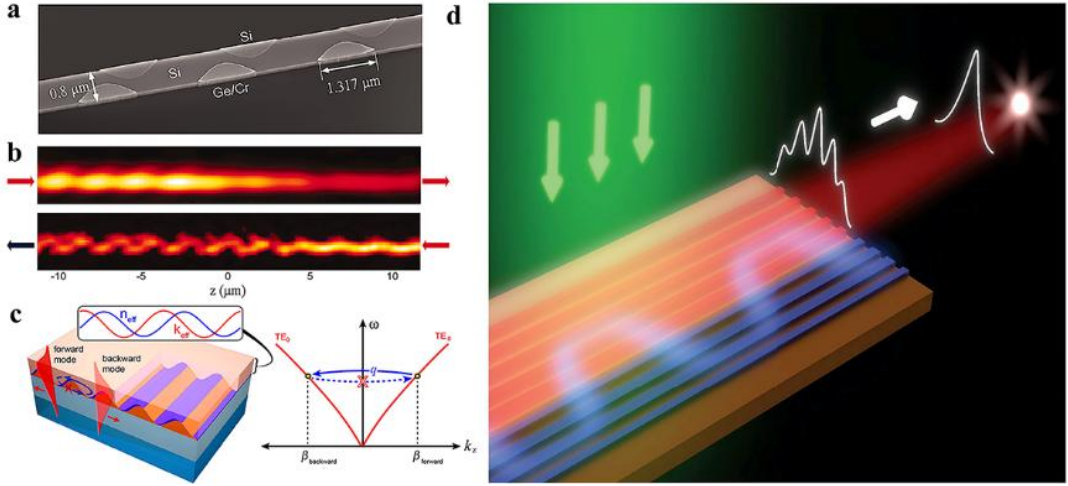

**Figure 9. Waveguide systems with non-Hermiticity and/PT symmetry. a)** SEM image of the fabricated waveguide. The thickness and width of the Si waveguide are 200 and 800 nm, respectively. The real and imaginary parts of the complex potentials are induced by 11 nm Ge/ 18 nm Cr and 40 nm additional Si. **b)** Experimentally observed near-field amplitude distribution for forward (upper) and backward (lower) waves of light. [48] **c)** Asymmetric behavior in a passive PT symmetric organic thin film metawaveguide. [95] The schematic diagram of the system with three layers (orange: photoresist, light turquoise: CYTOP (cyclized transparent optical polymer which is a low index fluoropolymer interlayer), dark turquoise: glass), showing propagation in forward and backward directions (left panel). Right panel shows the dispersion diagram of the system for fundamental TE mode. **d)** Schematic diagram of the SUSY-laser with active main array (red) and lossy superpartner array (blue).[94] Figures reproduced with permissions from: a,b) ref. [48] Copyright 2011, American Chemical Society; c) ref. [95] Copyright 2015, American Physical Society; d) ref. [94] Copyright 2019, American Association for the Advancement of Science.

More specifically it relies on the principle of unbroken SUSY for which the Hamiltonians $H^{(1,2)}$ are isospectral except the ground-state energy. Under proper pumping conditions, compared to single laser or conventional laser array, SUSY laser array reveals improved modal structure and lasing response, in addition to enhanced peak intensity (8.5 times the single laser and 4.2 times the conventional array).

## 3.2 Resonators

Resonator is one of the most widely used optical components, which has proved to be indispensable in achieving numerous intriguing effects. It possesses an inherent ability to trap light within a small region, thus, resonators can be treated as lossy media for certain wavelengths. It is also most widely used in the optical frequency comb generation largely due to its tunability via free spectral range or some frequency shifts. Despite its immense potentials, realization of actual microresonator-based PT systems has remained elusive. These integrated photonic structures could pioneer new investigations including topological diodes, enhancement of optical nonlinearity, lasing, sensing etc. where resonant structures and enhanced light-matter interactions are highly solicited.

Some experimental works demonstrated PT symmetry and PT phase transition in coupled whispering-gallery-mode (WGM) microcavities in linear regime, and nonreciprocal light transmission in presence of nonlinearity. [22,68] Two directly-coupled active-passive silica microtoroids (**Figure 10**a) were employed to study the PT breaking phase transition and gain saturation nonlinearity-induced nonreciprocal light transmission and optical isolation. [68] It was found that, when the coupling strengths ($\mu$) exceed the decay rates ($\gamma$) of the passive toroid ($\mu < \gamma$), the system resides in the unbroken PT phase with two spectral branches. In this unbroken PT phase, the output-input relation show linear response. At the critical point $\mu = \gamma$, the PT symmetry is spontaneously broken and coalescence occurs in the eigenmodes and the eigenspectrum. For $\mu < \gamma$, the system resides in the broken PT phase with the complex spectrum. In the broken PT phase, however, the output-input relation becomes nonlinear due to PT breaking-enhanced nonlinearity, which breaks the Lorentz reciprocity. **Figure 10**b,c show the transmission spectra for forward and backward light waves in different regimes of coupling and pump power conditions. In strong coupling regime ($\mu > \gamma$) with unbroken PT symmetry mode-splitting in the spectra can be seen with both low and high pump powers while in weak coupling regime ($\mu < \gamma$) with broken PT symmetry the two modes coalesce into an enhanced single mode for forward wave, which results in nonreciprocal isolation. In another work, in a system of two directly coupled whispering-gallery-mode resonators (WGMRs) with gain and loss, side-coupled with two tapered fibers (**Figure 10**d,e), [22] PT phase transition was observed as the inter-cavity coupling was varied. In addition, nonreciprocal transmission of light was demonstrated in the nonlinear regime with broken PT symmetry (see **Figure 10**f). In the absence of active gain elements, it shows bidirectional reciprocal response. In the presence of balanced gain and loss, it yields reciprocal output in the unbroken PT phase, while unidirectional nonreciprocal response in the broken PT phase. The nonreciprocal response in such systems stems from the enhanced gain saturation nonlinearity in the broken PT phase when increased field localization due to asymmetrical accumulation of a broken mode occurs in the active cavity. Thus, PT symmetric WGM-microresonators may be useful as excellent platforms for uncovering resonance phenomena in synthetic optical and photonic structures with versatile on-chip controllability.

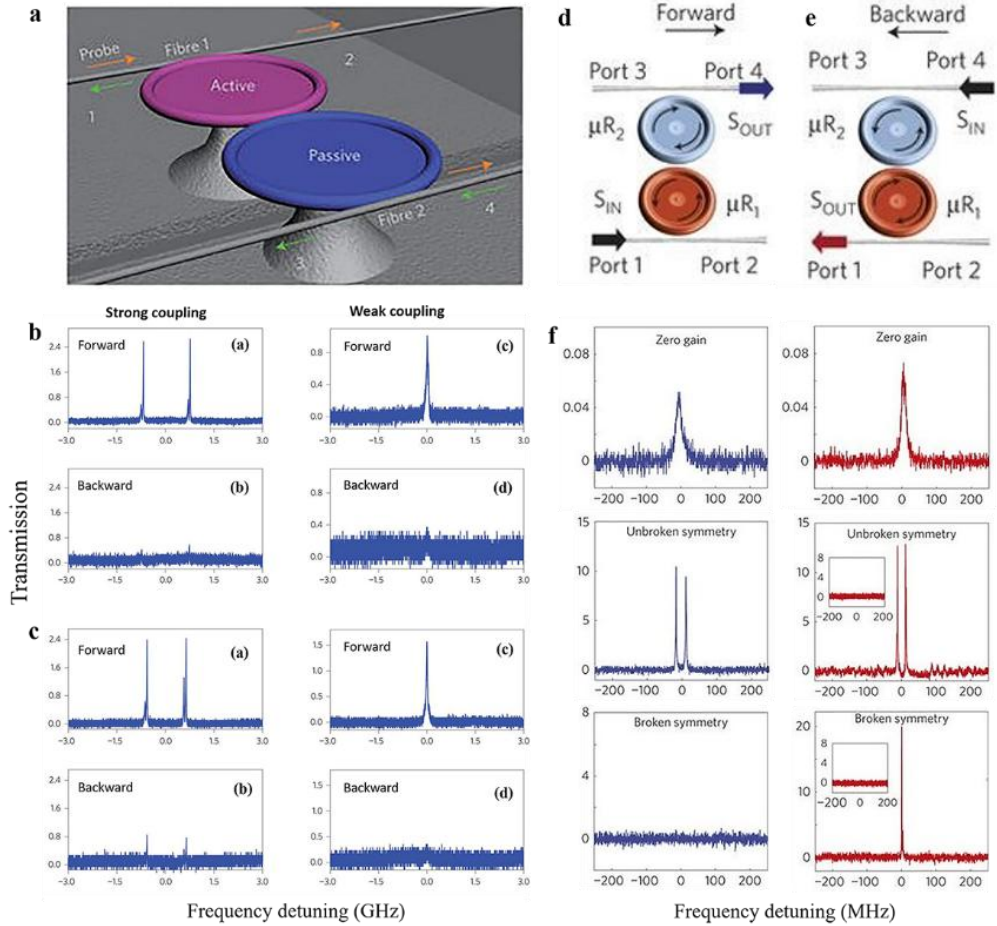

**Figure 10. PT symmetric resonators: coupled WGM microcavities. a)** Schematic representation of the PT symmetric resonators coupled to two tapered fibers. [68] **b,c)** Transmission spectra of the system versus the frequency detuning in different coupling regimes varying the coupling strength or the distance between the microtoroids of the system. For b) and c) input probe power is the same (11.4 nW) while they have different dropped pump powers (149.2 $\mu$W for b) and 149.2 $\mu$W for c)). **d-f)** The schematic illustrations of the PT symmetric microcavities side-coupled with the tapered fibers for (d) forward and (e) backward directions. (f) The unidirectional transmission spectra for forward (left panels) and backward (right panels) waves: passive (top), unbroken PT symmetry (middle), and broken PT symmetry (bottom). Insets: signal at port 1 when there is no input signal at port 4. [22] Figures reproduced with permissions from: a,b,c) ref. [68] Copyright 2014, Nature Publishing Group; d,e,f) ref. [22] Copyright 2014, Nature Publishing Group.

In demonstrating how the laser response is related to the EP physics, experimentally a coupled microdisk photonic molecule quantum cascade laser with active gain (**Figure 11**a) was studied in close proximity with the EP of the system. [53] The EP was induced in the vicinity of the first laser threshold where nonlinear effects are quite weak, and a linearized non-Hermitian Hamiltonian could effectively capture the inherent physics. Around the EP, it shows the characteristic reversal of the pump dependence behavior, for which the laser intensity decreases with the increase of the pump power. The lasing modes show frequency splitting as the pump power was swept (**Figure 11**b) when the lasing modes pass the EP. Such practical settings may be useful effective tools to investigate phenomena like CPA, [23,25] linewidth enhancement, [96] geometric phase accumulation in the vicinity of an EP. [97] In another work, PT symmetry breaking was exploited in realizing selective single-mode lasing in coupled microring resonators system of high refractive index contrast. [52] In such a typical laser setting, often the inhomogeneously broadened gain bandwidth gives rise to degrading competing modes in addition to the lasing mode. It was shown that selective PT breaking can be a useful tactic in strategically increasing amplification in the desired lasing mode, despite the presence of other competing modes in the amplification bandwidth, due to the fact the PT breaking solely depends of the coupling and gain/loss parameters. As a result, single longitudinal-mode lasing operation is achieved only for the balanced gain-loss PT resonators as shown in **Figure 11**c via judicious manipulation of the PT phase transition suppressing the undesirable competing modes. It is seen that single-mode lasing occurs in the gain resonator of the PT system ((c) of **Figure 11**c) when the pump power was employed in it as opposed to the cases where multi-mode response was found ((a), (b) of **Figure 11**c).

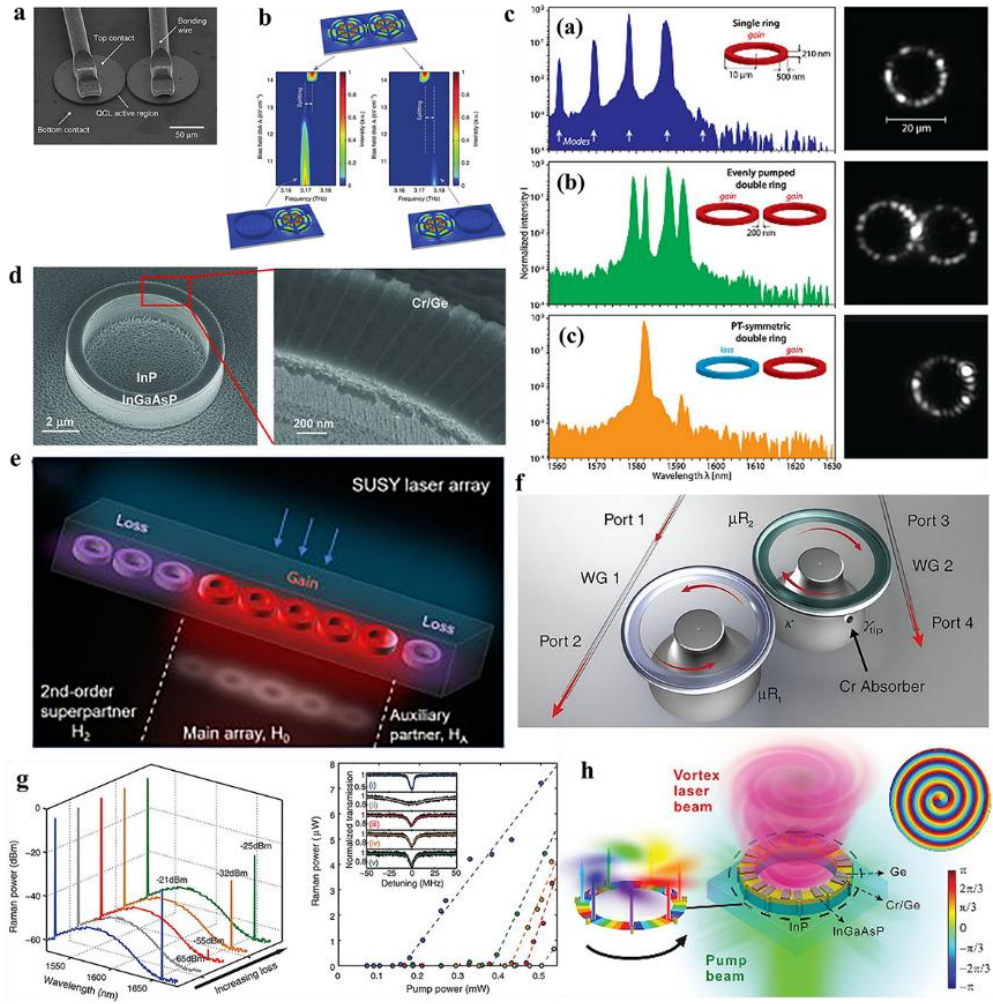

**Figure 11. Microresonator systems with non-Hermiticity and/PT symmetry. a)** The SEM image of a photonic molecule quantum cascade laser with disk radius 47$\mu$m, height 3.5 $\mu$m and intercavity distance 2 $\mu$m. **b)** Its measured emission spectrum vs the pump strength when pump light was used in the one cavity (left panel) with the pump kept at a fixed value slightly above the threshold value in the other or vice a versa (right panel). [53] The white dotted lines represent the frequency splitting as the lasing modes pass the EP. **c)** Different considerations of the gain-loss resonators: a singly pumped gain resonator (a), two equally pumped gain rings (b), and a PT symmetric pair of balanced gain-loss rings (c). Right panels: The intensity distribution patterns. [52] **d)** SEM images of the PT microring laser structure. [51] **e)** Schematic diagram of a non-Hermitian SUSY-based laser array is shown, depicting active main array (red), and superpartner array (purple). [99] **f)** The coupled microresonators and tapered fiber system to show the loss-induced effects on lasing. **g)** Raman lasing spectra of the system vs loss parameter (left panel). Laser output power vs the pump power to show the threshold of lasing response (right panel). Inset: The effect of loss on the threshold of lasing behavior. [98] **h)** Schematic illustration of the orbital angular momentum microlaser on InGaAsP/InP to achieve robust vortex lasing. [100] Figures reproduced with permissions from: a,b) ref. [53], Copyright 2014, Nature Publishing Group; c) ref. [52] Copyright 2014, American Association for the Advancement of Science; d) ref. [51] Copyright 2014, American Association for the Advancement of Science; e) ref. [99] Copyright

2019, Optical Society of America; f,g) ref. [98] Copyright 2014, American Association for the Advancement of Science; h) ref. [100] Copyright 2016, American Association for the Advancement of Science.

In experimental demonstration of single-mode lasing Feng *et al.* adopted a unique configuration of a single whispering-gallery-mode microring resonator. In this configuration, 500 nm thick InGaAsP multiple quantum wells were placed on top of InP substrate to form the microring structure. The gain-loss PT modulations were induced by the Cr-Ge structures on InGaAsP along the azimuthal direction (**Figure 11**d), which form the PT modulations. [51] The underlying continuous symmetry ensures the thresholdless PT breaking of the system for the specific WGM order modes, because of which all other modes stay below the lasing threshold except two energy-degenerate modes (one oscillatory gain mode and another lossy mode without oscillations). At lower level of pump power, broad photoluminescence emission spectra is observed while at higher pump power enhanced spontaneous emission and single-mode lasing around 1513 nm are exhibited. Remarkably, the PT microring system achieves single-mode lasing via selective lasing action on the specific WGM order lasing modes of its ordinary microring counterpart.

Concepts from symmetry paradigms such as SUSY can be useful tool in addressing practical issues as has been witnessed in recent times. Particularly in optics, it has promised novel possibilities in optical networks and laser arrays. Standard laser array systems suffer from multi-mode lasing spectra which can be exacerbated due to nonlinear mode competition among the different transverse lasing modes. It is found that appropriate loss-engineering of the laser array system along with ideas based on SUSY can significantly improve the lasing response and output beam quality. Midya *et al.* explores this possibility in a coupled microring system in which a standard laser array is coupled with its dissipative superpartner array with unbroken SUSY. [99]

Experimental results show enhanced single-mode lasing for SUSY laser compared to single or standard laser array (**Figure 11**e).

Another work shows how the degrading effects of optical loss can be addressed by converting loss into gain in the vicinity of an EP in a coupled microresonators system as shown in **Figure 12**f. [98] A critical value of loss was found below which addition of loss suppresses the lasing process, and beyond, the lasing recovers. As the overall loss was increased, the lasing response was suppressed and the lasing threshold increased (as is evident from the gray curve in **Figure 12**g). With further increase in loss, Raman lasing recovers and the lasing threshold decreases. On the other hand, in contrast to homogeneous planar light sources, orbital angular momentum (OAM) lasers promise significant potentials in optical communications and integrated photonics. To achieve the single-mode OAM vortex lasing in CMOS (complementary metal-dioxide-semiconductor) compatible platform, a WGM microring resonator with periodic gain-loss modulation in the azimuthal direction was considered (**Figure 12**h) that results into unidirectional lasing. [100] The single-mode vortex lasing was observed via index and gain-loss modulations at the EP with robust WGM modal selectivity. It shows that other less explored degrees of freedom of light such as OAM or polarization could add interesting attributes if properly harnessed.

The existence of the non-Hermitian degeneracy EP brings in many exotic effects not only for the fundamental understanding of the non-Hermitian physics, but also offers distinct application promises in many ways. It turns out that one can obtain highly enhanced sensitivity corresponding to environmental perturbations or fluctuations via a carefully engineered system operating at its EP. Compared to a Hermitian degeneracy point DP for which the frequency-splitting (measure of sensitivity) scales with external perturbation as $\Delta\omega \propto \varepsilon$, EP-assisted mechanism can yield the same as $\Delta\omega \propto \varepsilon^{1/N}$, $N$ indicates the order of degeneracy. A recent work experimentally explores such possibility due to a second-order EP in a silica microtoroid cavity coupled to a tapered fiber using two silica nanotips as Rayleigh scatterers. [101] As can be seen, **Figure 12**a,b show the linear Dirac-like eigenspectra and orthogonal eigenmodes for Hermitian (DP) degeneracy, while **Figure 12**c,d depict square-root topology and degenerate eigenmodes of the non-Hermitian point of degeneracy (EP). In the absence of the target scatterer, no frequency splitting is observed both for DP and EP (**Figure 12**e,g). However, introduction of the scatterer introduces enhanced frequency splitting and hence, sensitivity that stems from the joint contribution of the back-scattering from the target particle and the EP-assisted intrinsic back-scattering (**Figure 12**f,h). Another experimental work demonstrates similar enhanced EP sensing in a PT coupled microring-cavity system (**Figure 12**i-k). [55] The system exhibits a third-order EP and a cube-root dependence of the sensitivity on the perturbation. However, in a practical scenario, one needs to take into account the frequency variations of the individual resonators, or the temporal fluctuation associated with the gain or the frequency detuning, which may have adverse effect on the sensing by steering the system away from EP. [102-104] Another work shows fundamental bounds for the signal power and signal-to-noise ratio in such cases. [105]

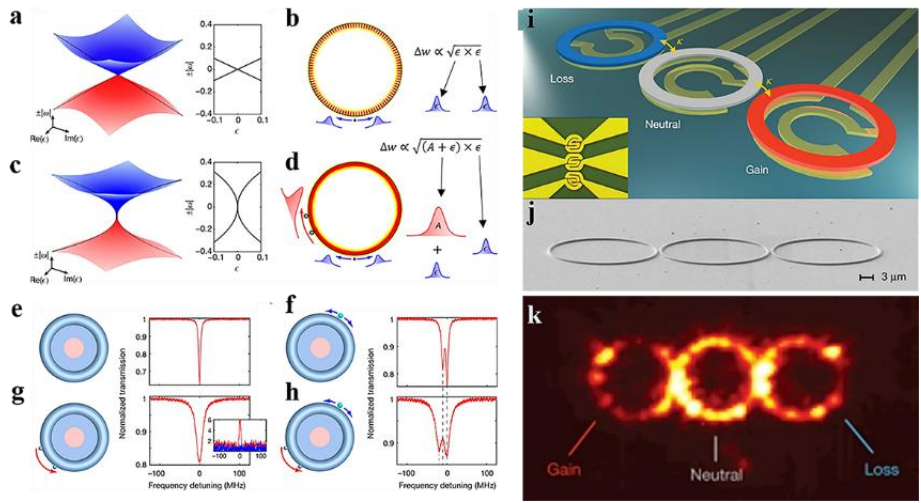

**Figure 12. Ultrasensitive sensing based on EP. a)** Linear eigenspectrum for Hermitian system containing a DP. **b)** The orthogonal eigenvectors corresponding to DP. **c)** Square-root eigenspectrum for non-Hermitian system containing a second-order EP. **d)** Coalescence of both eigenvalues and eigenvectors for EP. **e,f)** Transmission response for DP sensing before (**e**) and after (**f**) adsorption of the target scatterer. **g,h)** Transmission response for EP sensing before (**g**) and after (**h**) adsorption of the target scatterer. Inset in (**g**) shows asymmetric back-scattering from clock- (blue) and anti-clock-wise (red) mode of injection. The blue and red arrows in (**g**) indicate symmetric back-scattering from the scatterer and the intrinsic back-scattering enabled by the EP. [101] **i)** Schematic diagram of the PT trimer where the middle microcavity is without gain/loss and the adjacent sites are with balanced gain and loss. **j)** SEM image of the fabricated coupled microcavities. **k)** The intensity distribution of the lasing mode at a third order EP, which follows from the eigenmode of the system at EP, $|\varphi>_{EP3}=(1,-i\sqrt{2},-1)$. [55] Figures reproduced with permissions from: **a-h)** ref. [101] Copyright 2017, Springer Nature; **i-k)** ref. [55] Copyright 2017, Springer Nature.

**3.3 Photonic lattices**

Photonic lattices represent important class of platforms where one may find practical interests in view of their significant possibilities in optical communications and networks. Based on the types of configurations, we divide this section into three subsections based on coupled fiber loops, coupled photonic waveguides and bulk photonic lattices.

*3.3.1 Coupled fiber loops based photonic lattices*

Optical fiber is the most widely used optical medium which can guide light with very low propagation loss, enabling a long propagation distance and a large-scale networking capability. In quest of novel optical and photonic structures by proper manipulation of index, gain and loss modulations to harness the evolution and propagation dynamics of light to desired outcomes, a large-scale, temporal PT symmetric coupled fiber loop networks was proposed. [16] The condition of PT symmetry was established via alternately modulating gain and loss in the two fiber loops (**Figure 13**a,c). Light pulse evolutions in the short loops are advanced whereas those in the long loops delayed (**Figure 13**b). Unidirectional light propagation is observed in the scattering arrangement of PT symmetric temporal structure as depicted in **Figure 13**d. The passive version of the system shows usual broad scattering characteristics for incoming light from the left. For gain-loss modulated PT symmetric arrangement at the PT threshold, light is unidirectionally invisible with respect to the left (right) direction with reduced reflection (enhanced reflection). On the other side, in the first experimental study of the existence and dynamical properties of the defect states, [107] the defect was realized by including a pair of waveguides with different gain-loss contrast and/or phase shifts (**Figure 13**e). The transition of the stable and exponentially growing states have been exhibited with intact localization behaviors. Moreover, in the PT symmetric non-Hermitian lattices the defect states with complex spectra can reside within the band continuum.

In dissipative nonlinear systems one of the undesirable effects is uncontrollable growth of some of the eigenstates due to imaginary parts of the eigenvalues which together with nonlinearity leads to unpredictable behaviors. It was addressed experimentally via favorable interplay between nonlinearity and the gain-loss distributions as shown in **Figure 13**f-g. [108] In passive case, with single site excitation in the long loop, increasing input power gradually forms a single pulse dominated soliton switching between the two loops. In locally PT symmetric case, the same occurs with increasing input power, whereas in the global PT mesh lattice configuration, the existence of self-trapping nonlinear soliton states were found once the initial input power was increased. So, judicious interplay between nonlinearity and PT symmetry can witness stable nonlinear self-trapped states.

The practical design of many PT systems depends on the microstructures-based on-chip fabrication having large free-spectral range, which however, for larger macroscopic fiber network systems significantly decreases. So, whether the PT related phenomena still remain intact and robust in such networks poses a relevant question. In a recent experimental work, this particular issue was addressed in a system of more than one kilometer length. [106] It reports that despite the presence of detrimental effects of detuning between the subsystems of the whole network, the macroscopic system can yield characteristics of an EP, although exact coalescence of eigenstates does not occur due to statistical fluctuations.

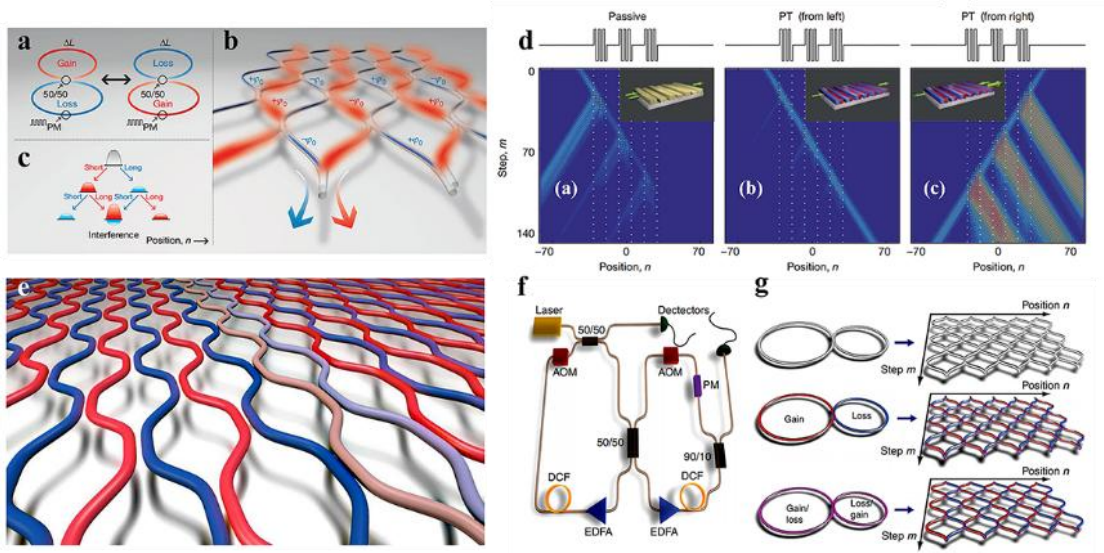

**Figure 13. Coupled fiber loops based PT symmetric photonic lattices. a)** Two coupled fiber loops with periodically modulated gain (red) and loss (blue) establishes the condition of PT symmetry. **b)** The PT symmetric coupled fiber loops network. **c)** Light pulse evolutions in the short loops are advanced whereas those in the long loops delayed. **d)** Experimental demonstration of the unidirectional reflection: scattering of a broad pulse ((a) for passive case); the invisibility of the gratings to a beam from the left ((b) at PT threshold); strong enhancement of reflection for right incident wave ((c)). [16] **e)** Illustration of a PT symmetric lattice network using temporal domain coupled fiber loops. Here, red (blue) waveguides indicate balanced regions of gain (loss) and a pair of waveguides denoted by light and dark gray colors act as defects. [107] **f)** Schematic diagram of the experimental setup is shown for realizing the discrete mesh lattices. [108] **g)** Different types of mesh lattices: Hermitian lattices (upper panel), lattices possessing local PT symmetry (middle panel), and globally PT symmetric lattices (bottom panel). [108] Figures reproduced with permissions from: a,b,c,d) ref. [16] Copyright 2012, Nature Publishing Group; e) ref. [107] Copyright 2013, American Physical Society; f,g) ref. [108] Copyright 2015, Nature Publishing Group.

*3.3.2 Waveguide based photonic lattices*

Waveguide-based photonic lattices constitute another important class of photonic systems. Due to initial explorations of PT concepts in coupled waveguides system, its immediate extension toward the waveguide lattices looked appealing.

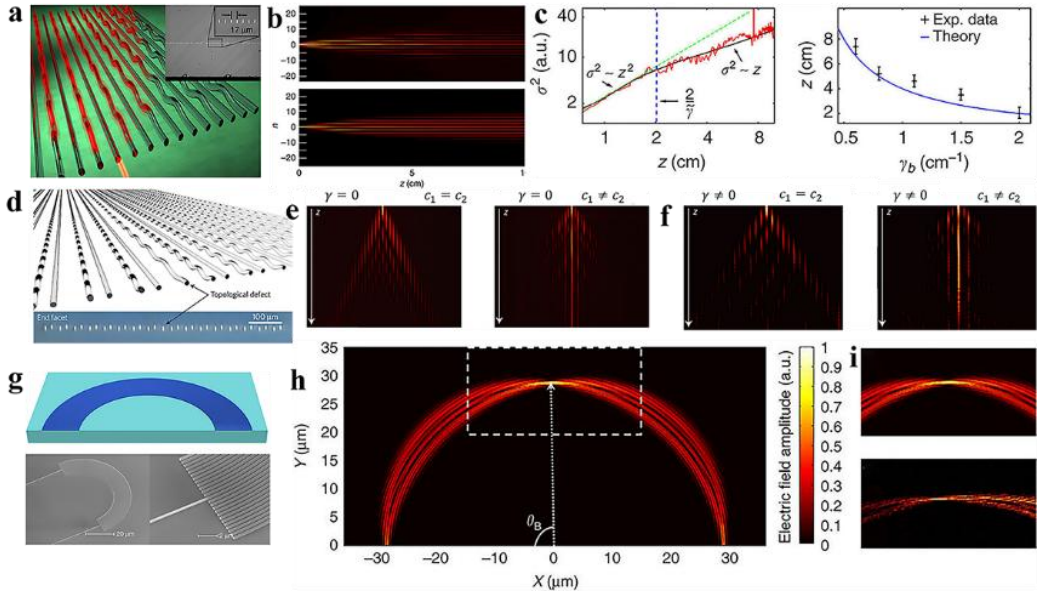

**Figure 14. Waveguide based PT symmetric photonic lattices. a)** Schematic diagram of the sinusoidally modulated dissipative waveguide arrays. Inset shows the microscopic image of the front side of the sample. [109] **b)** The propagation dynamics of light beam by Fluorescence microscopy in experiment (top) and simulation (bottom) has been shown. [109] **c)** The variance of the beam in experiment and simulation vs the propagation distance z (red line for experimental and black line for simulation results). The green and the blue dashed lines indicate the ballistic transport and the critical distance for the ballistic to diffusive response (left). The plot of the transition point vs the loss parameter $\gamma_b$ (right) is shown. [109] **d)** Schematic representation of the 1D passive PT symmetric structure with a topological defect at the center (top) is shown. The front facet of the experimental system (bottom) is demonstrated. [110] **e)** Experimental fluorescence images for lossless system: non-dimerized (left) and dimerized (right). [110] **f)** PT symmetric configuration with gain and loss: non-dimerized (left) and dimerized (right). Here $\gamma$ refers to gain/loss and $c_1$ and $c_2$ refer to coupling. [110] **g)** Schematic sketch of the non-Hermitian photonic lattice (top) is shown. SEM image of the fabricated photonic lattice and its zoom-in view is demonstrated. The central site denotes the input/output waveguide. [111] **h)** The electric field amplitudes distribution for single waveguide excitation in simulation is plotted. [111] **i)** Simulated (upper) and experimentally observed (lower) electric field distribution near the first BO recovery point (dash white box in (h)). [111] Figures reproduced with permissions from: a,b,c) ref. [109] Copyright 2013, Nature Publishing Group; d,e,f) ref. [110] Copyright 2016, Springer Nature; g,h,i) ref. [111] Copyright 2016, Nature Publishing Group.

It is believed that ballistic transport holds for ordered system, and disorder induces diffusion or localization behaviors in Hermitian context. In studying wave transport phenomena in dissipative non-Hermitian system, a coupled waveguides array was considered (**Figure 14**a).[109] The losses in such dissipative lattices were induced by sinusoidal modulations of the waveguides. It was theoretically predicted and experimentally demonstrated that in stark contrast to its Hermitian analogs, it exhibits initial ballistic spreading and modulated patterns, which after a critical propagation distance gets subdued (**Figure 14**b). This critical distance (see blue dashed line in **Figure 14**c) was calculated to be $z_{crit} = 2/\tilde{\gamma}$, ($\tilde{\gamma} = (\gamma_b - \gamma_a)$ is the gain/loss contrast), beyond which the ballistic to diffusive transition takes place. The ballistic to diffusive transition behavior was found to depend only on the dissipation of the system.

Topological photonics has opened a novel avenue of research due largely to the possibility of realizing robust transport phenomena, which together with non-Hermitian physics could induce novel effects. While some reports defied the possibility of robust topological edge states in non-Hermitian PT symmetric platforms due to PT symmetry breaking,[112-114] other proposals indicated quite the opposite.[115-118,110] Followed by realization of selective control and enhancement of topological edge states in 1D passive PT symmetric system,[118] topological edge states were realized with unbroken PT symmetry in an array of evanescently coupled waveguides with a defect waveguide at the center (see **Figure 14**d).[110] It was observed that dimerization leads to improved confinement of light in the defect waveguide (see **Figure 14**e) as it contains the interface state. On the other hand, loss in the non-dimerized case leads to broadening of light in the broken PT symmetry in sharp contrast to the dimerized case, where confined evolution of light in unbroken PT symmetry is observed (**Figure 14**f).

Bloch oscillation (BO) represents cyclical motion of particle under the action of a force in periodic media. Originally studied in solid state physics, BO has found potential ramification into diverse areas of physics such as cold atomic lattices, [119] waveguide arrays, [120] topological settings. [121] The observation of BO in photonic media is advantageous to electronic one due to availability of easy nanofabrication processes and accessibility of versatile photonic potentials. However, photonic systems entail to address various attenuation processes, which make them non-Hermitian in nature. It has been learned that judicious gain-loss tailoring of the photonic media can address the issue of loss and at the same time induce some interesting effects reminiscent of non-Hermiticity. [17, 122] In spatial domain, an experimental work demonstrates BO in non-Hermitian photonic CMOS system, (**Figure 14**g) in which the bending of the waveguides leads to a linear potential gradient. [111] In this system, 100 nm wide and 4 nm thick Cr layers were used to provide lossy components. The passive PT symmetric system results into thresholdless PT symmetry breaking under any excitation condition and the formation of complex Wannier-Stark ladders. The experimentally observed electric field distribution in the vicinity of the BO recovery point agreed well with the theoretical prediction (**Figure 14**h,i).

*3.3.3 Bulk Photonic crystal*

The bulk photonic crystals (PhCs) are PhCs with out of plane excitation waves, which provide yet another important class of systems for investigating peculiar effects due to non-Hermitian photonics.

To experimentally demonstrate the interconnection between Dirac and exceptional point, [86] interference photolithography was used to realize a periodic structure in $Si_3N_4$ on top of silica in a photonic crystal slab system (**Figure 15**a). In its Hermitian square lattice system, the variation of the system parameters leads to the accidental degeneracy between a quadruple mode and two degenerate dipole modes at the high symmetric $\Gamma$ point and the linear Dirac dispersion relation. Inclusion of gain-loss elements renders it to non-Hermitian system with open boundary condition. In this case, while the dipole mode still radiates, the quadruple mode does not contribute to radiation due to symmetry mismatch with the surrounding extended modes. In this open non-Hermitian system, the effective non-Hermitian Hamiltonian yields two complex eigenvalues. It exhibits a square-root branching behavior containing a 2D flatband within an exceptional ring in the k-space due to different radiation associated with the modes. It could be understood if we imagine the EP being adiabatically encircled varying a system parameter, the complex eigenvalues switch their positions along every k-space direction, forming the EP ring. Comparing the reflectivity spectra at three different angles in simulation and experiment, as shown in **Figure 15**b, it reveals that the reflectivity peaks are comparatively close to the actual positions of the complex eigenvalues in larger angles. In **Figure 15**c,d the theoretically predicted EP and the ring structure have been demonstrated experimentally. This work shows the possibility of exploring the topological synergy of the Dirac points and the exceptional points in practical photonic crystal settings, in addition to some interesting effects, such as photonic crystal lasers. More recently, the non-Hermiticity-induced bulk Fermi arcs and polarization half topological charges in a photonic crystal slab system exhibiting topological signatures have been demonstrated. [123] It was shown that a Dirac point splits into a pair of EPs when non-Hermitian effects from radiation-induced losses are taken into account. Remarkably, this EP pair is shown to possess the unique characteristics of a double-Riemann sheet topology, which gives rise to two intriguing results: the bulk Fermi arc in the form of an isofrequency contour that connects the EP pair and the topological half charges around the Fermi arc. In fact, the topological half charges prove the topological index $\nu = \pm 1/2$ of an EP, which describes the direction and number of winding around a singular point or line. Polarimetry results are obtained to properly reconstruct the far-field polarization profiles. As shown by the little green arrows in the numerical and experimental reconstruction results in **Figure 15**e-g, the polarization long-axis gives a winding of 180 degree as we start from the point X traversing the whole contour in the anti-clockwise direction and come back at X. This is due to the adiabatic mode-switching behavior of the EP along the closed contour. This work sheds light on the interconnection between Dirac and non-Hermitian degeneracies, and relates the polarization winding to the double-Riemann sheet topology of the paired EPs, and can in principle be important in unifying the band topology and the EP physics in non-Hermitian systems.

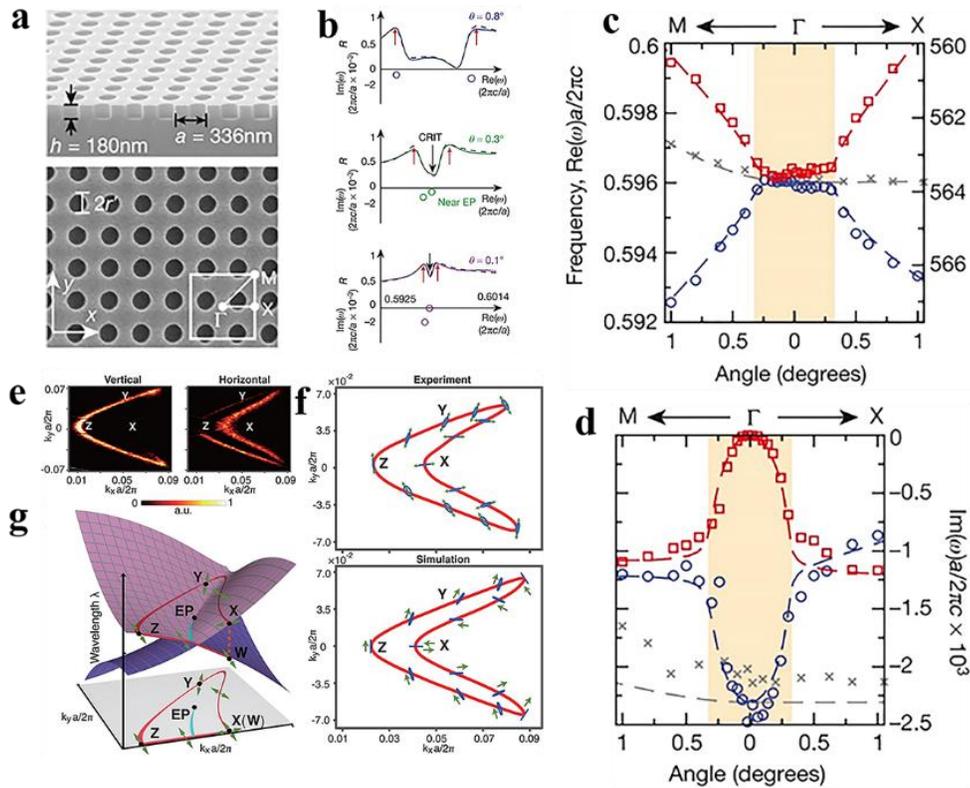

**Figure 15. Bulk photonic crystal structures: PhC slabs. a)** SEM images of the photonic crystal slab system showing the side (upper) and top view (lower). [86] **b)** The experimental reflection spectrum for different angles (0.8°, 0.3°, and 0.1°, denoted by blue, green and magenta solid lines; black line for the theoretical result). [86] Complex eigenvalues from experiment

(symbols), and numerical simulation (dash lines) for **c)** the real, **d)** and the imaginary (lower) part of the eigenvalues. **e-g)** The bulk Fermi arc and the topological half polarization charges in bulk PhC slab. [123] (e) The scattered light intensity in the vertical (left) and horizontal (right) polarizers. (f) The demonstration of the experimental (upper) and numerical (lower) polarization information along an isofrequency contour (red curve) as indicated by the blue ellipses and green arrows. (g) The mode switching behavior X→Y→Z→W along an isofrequency contour encircling an EP. Figures reproduced with permissions from: a,b,c,d) ref. [86] Copyright 2015, Nature Publishing Group; e,f,g) ref. [123] Copyright 2018, American Association for the Advancement of Science.

In an attempt to achieve reconfigurable PT system, Hahn *et al.* have adopted the idea of using a vector holographic interference technique in azo-dye-doped polymer thin films (**Figure 16**a). [65] Periodic pump optical fields $\boldsymbol{E}_p(x) = \boldsymbol{E}_1 + \boldsymbol{E}_2 = \boldsymbol{E}_p(x \pm \Lambda)$ induce surface-relief subgrating and the dichroic absorption subgrating structures which account for the real and imaginary parts of the dielectric functions respectively. Hence controlling the amplitudes and phases of $\Delta\varepsilon_R(x)$ and $\Delta\varepsilon_I(x)$ (dielectric function $\varepsilon(x) = \varepsilon_o + \Delta\varepsilon_R(x) + i\,\Delta\varepsilon_I(x)$) one can obtain expected non-Hermitian structure. Furthermore, in many experimental demonstrations of the electrically-injected coherently coupled semiconductor laser arrays, often the role of gain contrast or frequency detuning has been overshadowed by the current-controlled frequency or phase velocity tuning on the mode structures of the systems. In a vertical cavity surface emitting laser (VCSEL) arrays system, [124] the role of gain/loss contrast was studied experimentally. As shown in **Figure 16**b,c, with increasing $|\Delta I|$ one notices that the minimum intensity mode of the far field shifts toward the right, showing beam steering behavior.

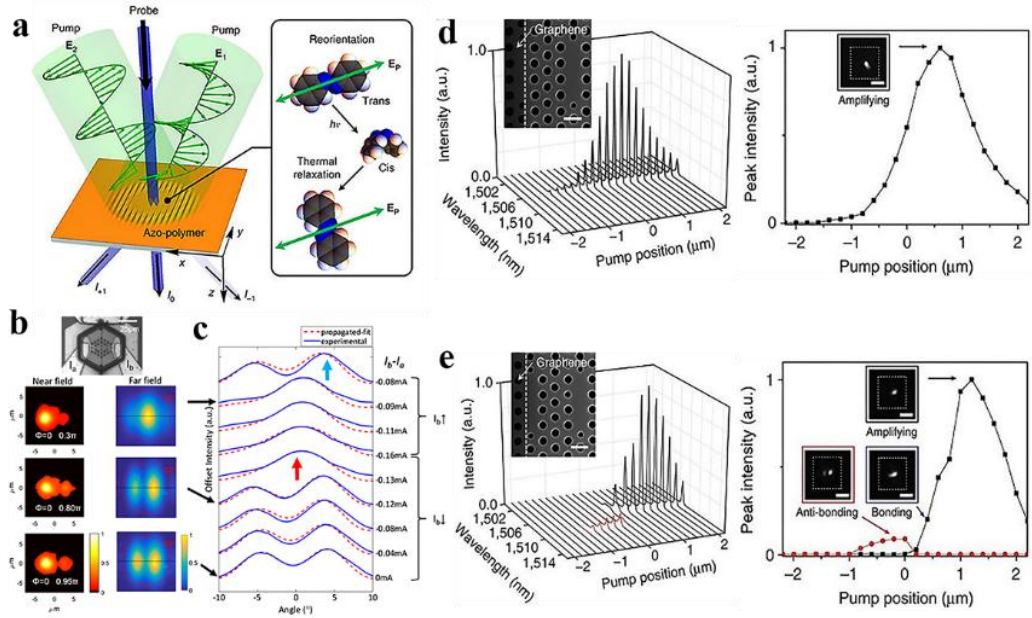

**Figure 16. Bulk photonic crystal structures: organic thin-films, VCSEL, graphene. a)** Illustrative diagram of the complex photonic lattice in the azo-dye-doped polymer thin films. [65] **b,c)** The left panel: the near-field intensity distributions. Insets: microscopic image of the vertical cavity surface emitting laser (VCSEL) diode array. The right panel: evolution of the far-field distributions. Here the solid blue line and the dash red line correspond to the experimental and propagated fit results. [124] **d,e)** The demonstrations of the scanning photoluminescence data in coupled photonic crystal cavities for large (first column) and small (second column) area graphene. [125] Figures reproduced with permissions from: a) ref. [65] Copyright 2016, Nature Publishing Group; b,c) ref. [124] Copyright 2016, Nature Publishing Group; d,e) ref. [125] Copyright 2017, Optical Society of America.

On the other side, the coupled cavity photonic structures have proved to be the playground of several counter-intuitive phenomena, especially for resonant lasing modes. [52,53,98] In conjunction with this, photonic crystal cavities with small mode volumes and high quality factors provide a class of indispensable platforms for non-Hermitian optical systems due to its capability of high mode selectivity and precise lasing response. Additionally, it has witnessed versatile possibilities such as, nanocavity laser, [126] symmetry breaking in coupled PhC nanolasers, [127] and Josephson interferometer. [128] In a non-Hermitian coupled optical cavities with asymmetric gain and identical loss, the complex eigenfrequencies are given by $w_\pm = w_0 + \sqrt{k^2 - \Delta\gamma^2} + i(\gamma_{avg} - K)$, where $w_0, k, K, \Delta\gamma, \gamma_{avg}$ are the eigenfrequency of the cavity, the coupling constant and the identical intrinsic loss of each cavity, gain contrast $(1/2|\gamma_2 - \gamma_1|)$ and average gain $(1/2(\gamma_2 + \gamma_1))$ respectively. The bonding and anti-bonding modes occur in the unbroken PT symmetry regime for small or identical gain-contrast ($k > \Delta\gamma$) of the cavities, whereas the decaying modes come into existence for the broken PT phase for larger gain-contrast ($k < \Delta\gamma$). As the asymmetric gain is provided to the cavities, the EP occurs when $k = \Delta\gamma$. The appearance of the amplifying or decaying modes corresponds to the condition $Im(w) > 0$ or $Im(w) < 0$ respectively. To observe these effects in experimental settings, closely coupled PhC cavities were fabricated on the InGaAsP photonic slab. [125] To provide proper gain-contrast and asymmetric gain in the cavities a single-layer of graphene was partially placed on top of one of the cavities. The effects of having a graphene sheet in one of the cavities on the lasing behaviors reveal that the presence and degree of loss induced by graphene critically decides the lasing response by enhancing gain-contrast of the cavities so that the split supermodes could coalesce into a single lasing mode in the broken PT phase. This single-mode lasing behavior (**Figure 16**d,e) originates from the amplifying supermode in the broken PT regime.

### 3.4 Hybrid photonic & other systems

In addition to the large number of optical and photonic systems as discussed earlier, the notions of PT symmetry have found due relevance into other potentially important platforms, including hybrid photonic systems such as exciton-polaritons, [129] and others such as metasurfaces, [88] microwave transmission line based absorbers,[67] quantum systems, [132] and so on.

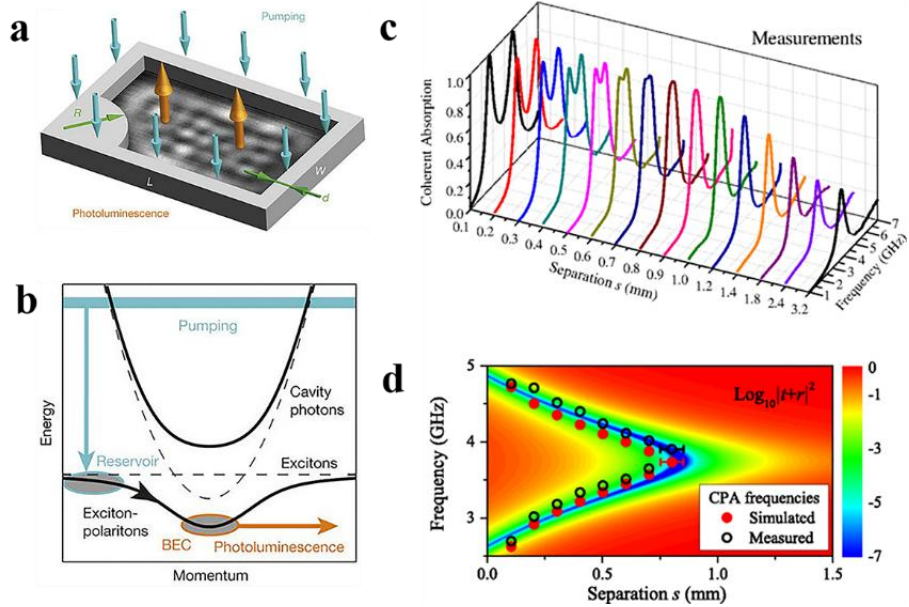

**Figure 17. Hybrid photonic and other PT symmetric systems: exciton-polaritons and microwave transmission line. a)** Schematic illustration of the non-Hermitian exciton-polaritons Sinai billiard system. Here, the billiard is located at one of the corners of the microcavity with radius 'R'. The cyan and the orange arrows represent the optical pumping and photoluminescence, respectively. [129] **b)** The dispersion plot of the exciton-polaritons system (a) is shown. **c)** Experimentally observed coherent perfect absorption as a function of the near-field coupling of the resonators. [67] **d)** Demonstration of the outgoing spectra in simulation (red solid) and measurement (black hollow) showing good agreement. Figures reproduced with permissions from: a,b) ref. [129] Copyright 2015, Nature Publishing Group; c,d) ref. [67] Copyright 2014, American Physical Society.

The hybrid quasiparticles exciton-polaritons are open quantum systems intrinsically representing the testing ground of non-Hermitian physics, and show collective quantum response in modal structures, spectral behaviors and dynamical properties even at room temperatures. They are also promising candidates of many intriguing effects in optoelectronic applications. The inherent gain-loss non-Hermitian attributes present in them naturally call for judicious exploration and manipulation. In the experiment, structured pump light was used to realize the non-Hermitian effects by forming a circular defect Sinai billiard with soft walls (see **Figure 17**a). [129] The optical pump (cavity photons) creates the excitonic reservoir with polaritons, which then hybridizes into BEC-like exciton-polaritons state. The BEC-like condensate emits photons upon relaxation process in the form of photoluminescence (**Figure 17**b). The variation in the radius of the Sinai billiard results in the changes in the energy levels and similar to the Hermitian case, multiple degeneracies and quasidegeneracies appear. But in sharp contrast to the Hermitian hard billiard case where the energy levels only exhibit anti-crossing, in this non-Hermitian soft billiard both level crossing and anti-crossing were observed.

After it was proposed, [130] in a subsequent work, [67] the ideal PT symmetric effects were practically realized in a passive system with electric and magnetic resonators coupled with a microwave transmission line. The coupled resonators system consider bright (made of copper) and dark resonators made of vertical wire and split-ring, respectively. Under appropriate choice of the parameter values and excitation conditions, ideal PT symmetry was shown to be established in the system. When the coupling between the resonators was varied, coherent perfect absorption was observed in the PT unbroken phase below a critical value in the coupling between the resonators, where it shows two CPA frequencies corresponding to the two modes, as depicted in **Figure 17** c,d. Above this value of the parameter, PT breaking occurs for which no CPA frequency can be found.

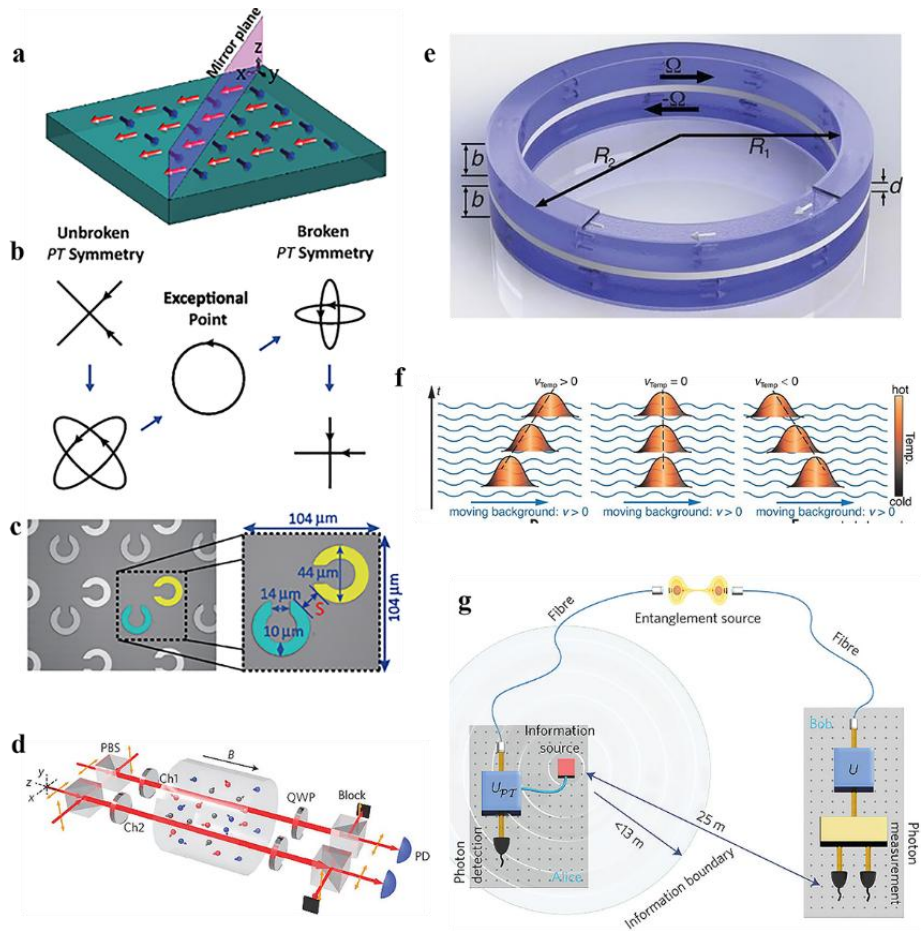

**Figure 18. Hybrid photonic and other PT symmetric systems: metamaterials, atomic media, thermodynamic and quantum systems. a)** Schematic diagram of the passive PT symmetric metasurface. Here red (blue) arrows indicate more lossy lead (less lossy silver) dipoles in terahertz domain. [88] **b)** Illustration of the different eigenstates of transmission in various PT phases. [88] **c)** PT symmetric metasurface with 300 nm thick silver (yellow or light gray) and lead (turquoise or dark gray) split ring resonators placed on silicon. [88] Inset: zoom-in view of the structure. **d)** The experimental setup for anti-PT symmetry in optics in an atomic-vapor system is demonstrated. [131] **e)** A 3D model of the diffusive heat transfer system is shown. $\Omega$ denotes the rotational speed. **f)** Schematic representation of different behaviors of the temperature profile of the system in various phases of anti-PT symmetry. [138] **g)** An experimental illustration of an open quantum system to show the no-signaling principle in PT symmetric theory is shown. [132] Figures reproduced with permissions from: a,b,c) ref. [88] Copyright 2014, American Physical Society; d) ref. [131] Copyright 2016, Nature Publishing Group; e,f) ref. [138] Copyright 2019, American Association for the Advancement of Science; g) ref. [132] Copyright 2016, Springer Nature.

Metamaterials provide unique versatile platforms for investigating many interesting effects in optics and photonics. Keeping in mind the immense potential of metamaterials, a 2D sheet of metasurface of coupled lossy resonators with two Lorentzian dipoles in perpendicular direction was considered (**Figure 18**a,c). [88] The PT phase transition was observed in the polarization eigenstates when the near-field coupling was varied by changing the coupling. For example, in the unbroken PT phase the polarization eigenstates of the Hamiltonian are also eigenstates of the PT operator and the eigenstates are represented by corotating ellipses with major axes inclined at 45 degree. In the broken phase, the polarization eigenstates are given by corotating ellipses at 0 and 90 degree. At the PT phase transition point, it yields only left circularly polarized light (**Figure 18**b).

In contrast to usual PT symmetry where $[\widehat{H}, \widehat{P}\widehat{T}] = 0$, the anti-PT symmetry $\{\widehat{H}, \widehat{P}\widehat{T}\} = 0$ can be useful in manipulating light in controllable way in non-Hermitian optics. Whereas the first theoretical work proposes a composite metamaterials system with balanced real and imaginary parts of the refractive indices, [133] subsequently a couple of investigations reveal that coherently prepared atomic systems can be ideal platforms for its practical implementation. [134-137] The coupling between two spatially separated probe fields via linear and nonlinear spin wave mixing is used to generate the non-Hermitian structure, as demonstrated in **Figure 18**d. On the other hand, a diffusive heat transfer system intrinsically represents a dissipative model, where non-Hermitian physics can be explored. On an interesting note, a recent experimental work establishes anti-PT symmetry in a dissipative heat convection system with two counterrotating rings (**Figure 18**e). It demonstrates spontaneous symmetry breaking phase transition from stationary (unbroken) to countermoving (broken) temperature profiles. [138] **Figure 18**f shows motionless (middle) and in-motion temperature profiles in opposite directions (left and right) w.r.t. the velocity of the background.

Despite the immense implications of PT physics in the classical systems, in quantum regime it results into some ideas that appear debatable to some, for example, quantum state discrimination, [139] ultrafast states transformation, [140] and no-signaling principle. [141] In particular, it was argued that as a local PT symmetric theory it violates the no-signaling principle. [141] In an experimental demonstration of the debated no-signaling principle violation, space-like separated two entangled photons were considered in which one of the entangled photons was described by the local PT symmetric formalism while the other one by the conventional quantum mechanics. [132] In the experimental setup as demonstrated in **Figure 18**g, the photon toward Alice is essentially simulated by a PT symmetric system by post-selected quantum gate. It was concluded that the violation of the no-signaling principle can be simulated in the experiment provided successfully evolved subspace with PT symmetry is taken into account.

## 4. Perspectives and future outlook

It could be noticed that starting from the initial interests in quantum mechanics and quantum field theories, the non-Hermitian complex extension of the existing laws and concepts have received significant research interests in the past two decades. In addition to the myriad of applications and possible areas of explorations, this emerging field is witnessing growing attentions in new areas with enhanced prospects for intriguing fundamental physics as well as engineering of novel materials. It includes prospective emerging areas, such as topological photonics, nonlinear and quantum optics, metamaterials and plasmonics, optomechanics, acoustics and so on. In the following, we provide extensive discussions on the emerging new directions of research along with future perspectives.

### 4.1 Non-Hermitian topological photonics

As an exciting class of solid-state materials, topological insulators provide unique platforms for exploring many fascinating ideas in photonic systems in analogy with its electronic counterparts. [142,143] In comparison to its electronic analog, topological ideas in photonic systems can benefit from a number of distinctive advantages. [144] Moreover, topological photonics can be an exciting platform for disorder-immune robust behaviors. On the other side, the gain-loss elements of non-Hermitian systems can enrich our generalized understanding of topological systems and add useful nontrivial design freedoms to the constructions of an extensive family of novel topological materials. The unlikelihood of the existence of the topological edge states in non-Hermitian systems, [112-114] later witnessed some positive results, [115-118,110] which in turn have spurred renewed interests. It was found that non-Hermiticity in combination with topology can yield intriguing behaviors otherwise absent in the Hermitian analogs, for example, the extraction of topological invariants via non-Hermitian effects. [113] More recently, demonstrating PT symmetric quantum walks with topological effects, [145] quantum dynamics of single photons was illustrated in a passive PT symmetric quantum system where the photon losses were stroboscopically modulated. The existence of topological edge states has been studied in non-Hermitian PT topological setting. [146,118] Recent theoretical, [147] and experimental, [148,149] explorations of topological lasers could trigger further interests in its complex non-Hermitian extension, for instance, topologically protected frequency comb active emitter. [150] Other potentially important areas could be PT symmetry protected topological semimetals, [151] photonic, [152] and acoustic, [232] topological systems, novel effects in photonic crystals, e.g. Weyl points, [153] Fermi arcs, [123] exceptional rings, [86] bulk and edge arcs, [154] and so on. In particular, as opposed to the electronic systems for realizing non-Abelian braiding of light for topological quantum computing, optical systems have been suggested as viable options owing largely to the all-optical integration capability. [155] However, one needs to address the robustness of the information under the unavoidable optical dissipation and losses where proper manipulation of non-Hermitian physics can be fruitful. Exploring topology in conjunction with weakly (mean-field Maxwell equations) or strongly (quantum optical settings) interacting nonlinear media could prove to be interesting. [156] Besides, non-Hermiticity can yield interesting effects in higher-order topological insulators. [157] For example, Zero-energy corner mode and chiral edge states are studied in non-Hermitian second-order topological insulators (SOTIs). [158] Moreover, one can think of unifying the topological and non-Hermitian degeneracies, [86,154,159] and symmetry protection of nodal and exceptional surfaces. [160,161] Recently, an experimental work shows emergent topological phenomena such as skyrmions via interplay between PT symmetry and quench dynamics. [162] On an interesting note, a more recent work demonstrates robust topological steering of light via non-Hermitian means. [163] Despite significant developments and future possibilities there may exist few challenges. These include the significance of bulk-boundary correspondence, [164,165] and the precise definition of topological invariants in non-Hermitian topological platforms, [166] the general framework constituting the basic tenets of non-Hermitian topological photonics. [167,168] It was realized that bulk-boundary correspondence does not hold, in general, for non-Hermitian Hamiltonians. In some of the interesting theoretical developments, some recent works attempt to address this particular issue in distinctive ways, for example, based on biorthogonal polarization, [169] non-Bloch topological invariants. [170] While the former idea based on biorthogonal polarization, generalizes the eigenbases for non-Hermitian Hamiltonians, [169] the latter proposes to introduce non-Bloch topological invariants that can predict the number of topological edge states. [170] More recently, another work correlates the breakdown of bulk-boundary correspondence in some non-Hermitian Hamiltonians to the absence of chiral-inversion symmetry protection. [171] It was claimed that judicious introduction of non-Hermiticity that preserves chiral-inversion symmetry, does not change the energy spectrum under open boundary conditions (OBC). The vortex and anti-vortex of the topological defects at the topological phase transition points give rise to the vorticity, the topological invariant of the system. In short, non-Hermitian attributes of topological systems necessitate introductions of new topological properties and modifications of the existing principles in the corresponding Hermitian counterparts. These theoretical predictions shed light on the controversial issue of bulk-boundary correspondence in non-Hermitian domain. It can trigger further theoretical and experimental interests in this direction toward a unified coherent understanding of the longstanding problem. In addition, symmetry and topology can intertwine to yield intriguing effects in generalized non-Hermitian settings. [172,173] It can be speculated that such topological classification of non-Hermitian systems based on symmetries can lead to further developments in non-Hermitian topological photonics with gain and/or loss. Moreover, it was argued that pseudo-Hermiticity can provide nontrivial topological structure. So, non-Hermitian topological systems can characteristically induce new topological properties and modifications in the existing Hermitian analogs, which if properly harnessed, may possess future potentials for novel topological materials.

**4.2 PT symmetry in Nonlinear optics**

The interplay of nonlinear effects and non-Hermitian PT symmetric theory could be an interesting field of active research in future. This could be due to the naturally coexisting nature of nonlinearity and non-Hermitian gain/loss elements, which calls for immediate explorations. The existence, stability and dynamics of different continuous, [174-178] and discrete, [179,180] families of solitons pose an interesting aspect in PT symmetric nonlinear systems where nonlinearity provides an extra degree of freedom for phase transition and wave manipulation. As opposed to the Ginzburg-Landau equations used for modeling dissipative solitons, PT symmetry offers parametric flexibility, stability and yields families of soliton solutions. On the other hand, in contrast to the standard NLSE models, where only bright or dark solitons are found, PT symmetric nonlinearity results in simultaneous bright and dark solitons, [181] and Peregrine rogue waves. [182] On a more fundamental level, it could be interesting to take a note of some recent works studying nonlinearity in combination with non-Hermiticity, for example, solitary waves and peculiar new effects in PT symmetric nonlinear Dirac equation (PT-NLDE), [183] 3D topological solitons in complex PT lattices. [184] Furthermore, nonlinearity in conjunction with PT related ideas sheds light on nonreciprocal behavior in coupled WGM microtoroids systems and one can expect it to trigger further investigations of intriguing resonant phenomena in such synthetic optical and photonic structures. [22,68] Nonlinear effects in non-Hermitian PT symmetric coupled waveguides can result in unidirectional response, above a critical value of strength of nonlinearity, [185] which can find useful exploration in directed wave transport in nonlinear devices. In addition, in contrast to the usual PT phase transition via a non-Hermitian degenerate EP, PT symmetry breaking was found to occur away from the EP with the increase of the nonlinearity. [186] PT symmetry-induced effects can be interesting in nonlinear switching of optical signals. [85,187] On the other side, lasers could provide important class of platforms to investigate nonlinear effects along with non-Hermitian physics, above the first lasing threshold, where lasers are nonlinear in nature. In this case, nonlinear modal interactions may play an important role. [83,188]

**4.3 PT symmetry in quantum domain**

Optical systems involving gain and loss should be treated as open quantum systems with the dynamics described by the Lindblad-type equations, [189,190] or through Naimark's dilation. [191,192] In the macroscopic limit, the effect of noise may be ignored and a non-Hermitian mean-field description can be applied. However, in the microscopic region the non-Hermitian quantum mechanics possesses many striking features compared to the conventional quantum mechanics. For example, the quantum state may evolve much faster with non-Hermitian PT symmetric Hamiltonian, [140] and even violate the no-signaling principle. [141,132] How to interpret these results is still under debate. It is worth mentioning that quantum noises or fluctuations are related to gain and loss mechanisms through the fluctuation-dissipation theorem, which may indicate plausible synergy. Quantum noise has been studied in PT symmetric scattering system in which sustained radiation originated in the leaky condition. [193] There have been efforts to provide a complete description of the non-Hermitian PT symmetric systems using the conventional quantum mechanics. [194-196] Recently, there have been few notable developments in this direction, for instance, experimental investigation of no-signaling principle using space-like separated entangled photons, [132] the effects of complex PT symmetric potentials on the quantum transport were studied in a coupled double quantum-dot structure due to modified quantum interference, [197] possibility of robust coherence in PT symmetric qubits. [198] Another work theoretically proposes a circuit-QED structure in which the dynamics was modeled by an effective non-Hermitian Hamiltonian with PT symmetry. [199] The system consists of two coupled microwave cavities with a qubit placed near each of the cavities. A critical value of the coupling between the resonators was found that corresponds to the PT breaking EP. Of interest will be to investigate quantum interference or correlation in PT systems and the effects due to nonlinear interactions. Besides, cavity quantum electrodynamics, [200] driven atomic system, [201] and optomechanics (see following discussion) can be some of the quantum related platforms to explore notions of PT symmetry in quantum domain. Recent experimental demonstration of PT symmetry breaking exceptional point in single spin-based quantum systems may be promising in harnessing PT related ideas in a quantum sensing. [202] Some other related works may include, strongly correlated quantum critical effects in combination with PT singularity, [203] extended quantum geometric tensor to tune the PT symmetric systems at the quantum criticality. [204] More recently, two particle quantum interference has been experimentally demonstrated in a PT symmetric photonic waveguides system. [205]

Despite the fact that it has not been explored extensively and more studies are needed, PT symmetric quantum mechanics may enable novel applications in quantum information processing.

**4.4 Optomechanics**

Notable advances have been achieved in the field of cavity-optomechanics, where interaction between the electromagnetic and mechanical systems are manipulated to observe some of the interesting physical effects, for example, optomechanically-induced transparency (OMIT), phonon lasers, sensing, chaotic dynamics, wavelength conversion and so on. [206-209] Enhanced quantum nonlinearities and optomechanical interactions in such systems provide unique possibilities and versatile platforms for novel effects and functionalities. Additionally, non-Hermitian physics based on PT symmetry can add another nontrivial aspect in such settings. For example, strong nonlinear response and phonon lasing have been observed near the PT phase transition point (EP) in a gain-loss balanced optomechanical microcavities. [210] PT symmetry breaking chaos was found in an optomechanical system with ultralow power threshold. [211] This could be potentially important not only for manipulating chaos-assisted applications, but also bringing nonlinear optics and emerging field of PT optomechanics together at the forefront of the far-reaching possibilities. Optomechanically-induced transparency (OMIT) has been reported in PT symmetric optomechanical microcavities, [212-215] which promises new possibilities for EP physics and optomechanics. Some other important developments may include higher-order EPs for high-precision sensing, [216] loss-induced transparency. [217] It can be speculated that judicious exploitation and synergy of non-Hermitian physics and cavity optomechanics could shed new light in classical or quantum regimes.

**4.5 Metamaterials and Plasmonics**

The electromagnetic constitutive quantities are now accessible in the entire complex plane, where the imaginary parts correspond to gain or loss. The spatial distributions of gain and loss can not only be viewed as viable means to compensate detrimental effects of loss, but also induce some of the unconventional effects in the broad spectrum of platforms for light-matter interaction. As artificially engineered subwavelength structured media metamaterials can offer a unique set of platforms for hosting physically intriguing behaviors in non-Hermitian domain, e.g. unidirectional reflectionless metamaterial at optical frequency, [21] and in plasmonic waveguides, [90] PT singularity-enhanced sensing, [218] PT symmetry breaking, [67,88,130] coherent perfect absorbers. [67] On the other hand, once realization of aberration-free and transversely invariant ideal lens has been theorized, [219] great deal of research interests bloomed around double-negative (DNG) lenses when soon it was understood that the ideal DNG lens in fact gives rise to reduced longitudinal resolution. Other possible solutions in terms of nonlinear surfaces and photonic crystals fibers were found to be limited. Recently, using the scattering properties of PT symmetric metasurfaces, a loss-immune, all-angle highly efficient negative refraction has been studied with potential volumetric imaging ameliorating some of the preexisting limitations and widening further scope of ideal optical imaging. [220,221] Moreover, in order to efficiently describe the wave manipulation and waveguiding properties of non-Hermitian metamaterials, complex-coordinate transformation optics has been formulated. [222] On the other side, the dispersion and dissipation are less tunable in naturally-occurring media. In metamaterials they provide tunability over a rather limited frequency range. Non-Hermitian metamaterials as structured platform can provide useful means toward this aim based on the Kramers-Kronig (K-K) relations by appropriate design of dispersion via engineered dissipation. [223] In addition, distinct features of spatial K-K relations in artificial metamaterials can stimulate further interests in non-Hermitian systems. [74]

In a different direction, the field of plasmonics endeavors to achieve controllability and enhanced localization of light. In both types of surface plasmons, surface plasmon polaritons (SPPs), [224] and localized surface plasmons (LSPs), [225] non-Hermitian physics related ideas are likely to provide another useful degree of freedom to manipulate dissipation. Plasmonic systems can be utilized in useful device engineering while manipulating light localization in combination with the notions of PT symmetry. [89,90] Some recent works propose to exploit PT symmetric concepts in plasmonic devices, for example, active polarization control of light using coaxial PT nanoplasmonic structure, [226] multiplexing in coaxial plasmonic waveguides. [227] Despite the significant possibilities of PT related ideas in plasmonics, its practical implementation requires precise manipulation in the distributions of the gain-loss elements at subwavelength scale, which might be difficult at times. So, accessing EP physics in general plasmonic systems can be challenging. It will be interesting to study EP related effects in plasmonic systems, specifically in experimental settings.

**4.6 PT acoustics**

Due to similarity between the governing principles, exploring PT symmetry related ideas in acoustics could prove to be useful, especially keeping in mind the potential advancements of acoustic metamaterials including, asymmetric acoustic transmission, [228,229] subwavelength imaging, [230] acoustic cloaking, [231] acoustic topological insulator, [232] and so on. In acoustic systems the gain-loss distributions can be induced via several ways, for example, using loudspeakers loaded with electronic circuits to absorb or inject energy, [233-235] controlling the electric bias in piezoelectric semiconductor slab. [236] In recent times, PT symmetry has been explored in different acoustic systems, e.g. PT symmetry breaking in airflow duct, [237] the emergence and interaction of multiple EPs in coupled acoustic systems, [238] constant-pressure acoustic wave in a disordered non-Hermitian metamaterials, [239] enhanced linewidth broadening of a phonon laser in an optomechanical coupled resonators at an EP. [240] More recently, how PT symmetric gain-loss distributions can yield new localization rules for sound waves, has been demonstrated in a second-order topological sonic crystal. [241] Thus, it can be expected that phononic crystals would represent versatile platforms to explore rich physics of non-Hermitian PT symmetric systems.

## 4.7 Other symmetry paradigms

Symmetries are ubiquitous in nature and they dictate conservation laws. Symmetry paradigms, closely related with PT symmetry, can result in altogether new physical effects and possibility for practical utilization. In contrast to PT symmetry, its rotational variant RT symmetry can give rise to altogether different qualitative behaviors, [242,243] for example, the spectral transition requires finite non-Hermitian strength in RT symmetry whereas it can be finite, infinitesimal or both in PT. [243] Even interesting is the proposition that in absence of stricter condition of balanced gain-loss distribution, one can still obtain a mode with real propagation constant and PT-like response in dissimilar couplers. [244] On the other hand, supersymmetry (SUSY) in particle physics attempts to unify all the fundamental interactions of nature. From particle physics and quantum field theory, it has spread into optics, [245,246] condensed matter, [247] and it has been utilized in coupled optical networks, [245,248] transformation optics, [246] and laser arrays. [249] Despite the fact that its credibility as a unifying theory still remains to be validated in experiments, it has witnessed widespread ramification in several other areas of physics. In optics too it has observed some notable progresses recently. [250,251] It could be interesting to explore SUSY related ideas in non-Hermitian contexts, where the interplay between non-Hermitian physics and SUSY formalism can develop new strategies in practical settings. Some recent demonstrations of loss-engineered SUSY-based laser with single-mode operation and enhanced intensity show its potential application. [99,94,252] Along similar lines, particle-hole symmetry is shown to be a general attribute of non-Hermitian photonic configurations with two sublattices, which results in symmetry protection of the zero modes leading to a tunable and single-mode laser. [253] In fact, anti-PT symmetry, [254] for which the Hamiltonian of the system anticommutes with the combined PT operator, is a special case of non-Hermitian particle-Hole symmetry (NHPH). A recent work reveals topological equivalence of time-reversal and NHPH symmetries in the language of non-Hermitian physics. [255] Moreover, symmetry protection of nodal and exceptional surfaces has been discussed due to PT and NHPH symmetries. [160,161] Besides, non-Hermitian chiral symmetry (NHCS) has given rise to interesting effects, such as quantized complex Berry phase, [256] recovery of bulk-boundary correspondence, [171] symmetry-protected exceptional rings. [257] On the other hand, anti-PT symmetry shows significant promises as is evident from some of the recent developments such as: coherent switch in birefringent waveguides, [258] energy-difference conserving dynamics and reverse PT transition in electrical circuits, [259] constant refraction in dissipatively coupled system, [260] experimental realization of anti-PT symmetry in atomic medium. [131] More recently, anti-PT symmetry-induced unique effects were studied in diffusive heat convection system where the temperature profile witnesses stationary behavior in unbroken phase, and movement along or against the flow in the broken phase. [138] Moreover, Majorana zero modes have been considered a prominent candidate for realizing robust topological quantum computing largely due to their non-Abelian and topological characteristics. [261-263] There are instances of photonic robust zero modes in non-Hermitian systems which might bear some nontrivial connections. [264-266] Moreover, distinct features of quasiparticles are found in PT symmetric superconducting model within a mean-field approach. [267] Another recent work explores the unconventional Majorana edge modes in an ideal gain-loss PT symmetric topological superconducting wire with nonlocal anomalous current transport at the edges due to interplay between PT symmetry and topology. [268] It can be interesting to look into the possibility of exploring non-Hermitian PT physics in topological quantum computing, where PT related ideas may induce some novel effects.

While state of the art theoretical and experimental methods and techniques have already resulted into many significant and breakthrough achievements in fundamental physics and novel materials in non-Hermitian physics based on PT symmetry, new and efficient powerful tools can enhance the future possibilities. Some powerful approaches could be pursued in exploring some of the intricate physical effects reminiscent of PT optical and photonic structures, for instance, group theoretical method in continuous and discrete systems. [269] Recent discovery of a novel technique promises ultralow loss fabrication of lithium niobate-based small-scale optical microstructures, which can widen the future scope in efficient integrated photonic circuits, including PT symmetry based optical, photonic and opto-electronic systems. [270]

**4.8 Toward Functional devices**

One can easily note that non-Hermitian physics and PT symmetry related ideas from initial interests in quantum field theories have permeated a diverse range of subjects in fundamental as well as applied territories. The theoretical predictions were subsequently exhibited in a number of pioneering experimental demonstrations. It has unveiled a myriad of intriguing physical effects, which are fundamentally stimulating and technologically appealing. It can be speculated that much of the theoretically predicted and experimentally realized behaviors pave the way toward novel functional components and materials. Examples may include, efficient single-mode lasers, high-power sources, and ultrasensitive sensors and so on. However, practical utilization of the PT symmetry related devices may entail careful engineering of the systems and an overall consideration of effects pertaining to the systems or environments could be important. [102, 271]

**5. Conclusion**

A state of the art account of the recent developments in a wide range of areas involving the emerging idea PT symmetry in complex non-Hermitian media has been presented along with brief background and the basic physical and mathematical formulations. It has been noted that since its inception and initial interests in quantum mechanics and quantum field theory, it has rapidly spread into a diverse set of areas with significant cross-disciplinary implications both in terms of fundamental physics and novel applications. In particular, it has received immense activity and research interests in optics and photonics due to formal equivalence between the governing mathematical equations. This way it has provided a viable option to exploit the detrimental effects of loss and attenuation present in the systems via appropriate loss engineering or gain-loss modulations not only to induce some of the highly desirable physical effects and phenomena but also ensure enhanced controllability and versatile manipulation in a wide range of physical systems. It is for this reason that this emerging field can prove to be pivotal in the engineering of advanced functional components and materials in many areas of physics and engineering, including optics and photonics. In addition, a vast range of the newly emerging fields can also ensure a cross-disciplinary synergy and explorations of the existing and new ideas. It can be expected, especially at the advent of new and efficient nanofabrication and quantum technologies, that in the near future we can witness its far reaching impacts in much more inclusive and cross-disciplinary way for the next-generation of intriguing physical effects and devices in optics, photonics, opto-electronics, condensed matter and beyond.

**Conflict of Interests**

The authors declare no conflict of interests.

**Acknowledgments**

This work was supported by the National Key R&D Program of China (2017YFA0303702 and 2015CB659400), National Natural Science Foundation of China (Grant No.11625418, No. 11474158, No. 51732006, No. 51721001, Grant No. 11690032, No. 11474159 and No. 51472114), and the Natural Science Foundation of Jiangsu Province (BK20140019). L.Z. acknowledges funding support from the National Natural Science Foundation of China (Grant No. 11690032, No. 61490711, No. 91536113, and No. 11474159.

**Glossary:**

PT: Parity-time
NH: Non-Hermitian
PTQM: Parity-time symmetric quantum mechanics
NHPT: Non-Hermitian physics and PT symmetry
EP: Exceptional point
DP: Diabolical point
PED: Paraxial equation of diffraction
SE: Schrödinger equation
NLSE: Nonlinear Schrödinger equation
CPA: Coherent perfect absorber/absorption
CPT: Charge, parity, and time-reversal
CMOS: Complementary metal-dioxide-semiconductor
SOI: Silicon-on-insulator
WGM: Whispering-gallery mode
WGMR: Whispering-gallery mode resonator
OAM: Orbital angular momentum
SEM: Scanning electron microscope
1D: One dimension/dimensional
2D: Two dimension/dimensional
3D: Three dimension/dimensional
BO: Bloch oscillation
PhC: Photonic crystal
VCSEL: Vertical cavity surface emitting laser
BEC: Bose-Einstein condensate
QED: Quantum-electrodynamics
SOTI: Second-order topological insulator
OBC: Open boundary condition
PT-NLDE: Parity-time symmetric nonlinear Dirac equation
OMIT: Optomechanically-induced transparency
DNG: Double-negative
K-K: Kramers-Kronig
RT: Rotation-time
SUSY: Supersymmetry
NHPH: Non-Hermitian particle-hole
NHCS: Non-Hermitian chiral symmetry

**Biographies:**

**Dr. Samit Kumar Gupta** received his Ph.D. in 2016 from the Dept. of Physics, Indian Institute of Technology Guwahati, India. Afterward, in 2017 he joined the College of Engineering and Applied Sciences, National Laboratory of Solid State Microstructures, Nanjing University as a Postdoc Fellow. His research interests include fundamental and applied aspects of nonlinear optics, nonlinear waves, non-Hermitian physics, and topological photonics.

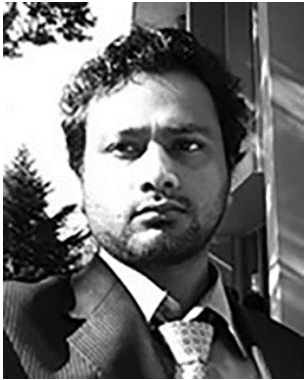

**Prof. Ming-Hui Lu** received his Ph.D. degree from Nanjing University in 2007. He is an Associate Professor at Nanjing University since 2009 and a Professor in 2013. He had been a visiting scholar at SIMES, Stanford University during 2011-12. His current research interests mainly focus on fundamental study of photonic and acoustic artificial structures and metamaterials as well as their related applications.

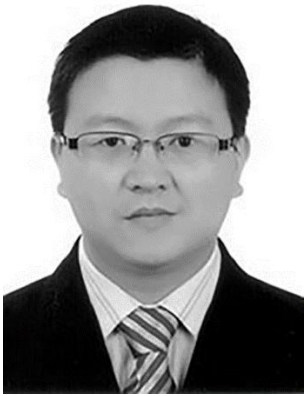

**Prof. Yan-Feng Chen** received his Ph.D. degree in 1990 in materials science and engineering from Northwestern Polytechnic University, Xi'an, China. He was a Postdoc Fellow at Dept. of Physics, Nanjing University (1990-93) and a visiting scholar at Dept. of Physics, HKUST, Hong Kong (1998-99) and Dept. of Materials Science and Engineering at MIT, USA (2000-01). He became an Associate Professor (1993), and a Full Professor (1997) at Dept. of Materials Science and Engineering, Nanjing University. His research interests focus on phononic and photonic crystals, metamaterials, and superlattices.

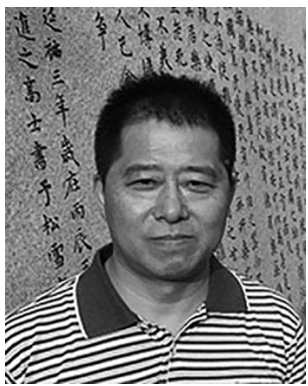